# Electron ptychography achieves atomic-resolution limits set by lattice vibrations

Zhen Chen[1*], Yi Jiang[2], Yu-Tsun Shao[1], Megan E. Holtz[3], Michal Odstrčil[4†], Manuel Guizar-Sicairos[4], Isabelle Hanke[5], Steffen Ganschow[5], Darrell G. Schlom[3,5,6], David A. Muller[1,6*]

[1]School of Applied and Engineering Physics, Cornell University, Ithaca, NY 14853, USA

[2]Advanced Photon Source, Argonne National Laboratory, Lemont, IL 60439, USA

[3]Department of Materials Science and Engineering, Cornell University, Ithaca, NY, USA

[4]Paul Scherrer Institut, 5232 Villigen PSI, Switzerland

[5]Leibniz-Institut für Kristallzüchtung, Max-Born-Str. 2, 12489 Berlin, Germany

[6]Kavli Institute at Cornell for Nanoscale Science, Ithaca, NY, USA

* Correspondence to: zhen.chen@cornell.edu (Z.C.); david.a.muller@cornell.edu (D.A.M.)

†Present address: Carl Zeiss SMT, Carl-Zeiss-Straße 22, 73447 Oberkochen, Germany

**Abstract:**

Transmission electron microscopes use electrons with wavelengths of a few picometers, potentially capable of imaging individual atoms in solids at a resolution ultimately set by the intrinsic size of an atom. Unfortunately, due to imperfections in the imaging lenses and multiple scattering of electrons in the sample, the image resolution reached is 3 to 10 times worse. Here, by inversely solving the multiple scattering problem and overcoming the aberrations of the electron probe using electron ptychography to recover a linear phase response in thick samples, we demonstrate an instrumental blurring of under 20 picometers. The widths of atomic columns in the measured electrostatic potential are now no longer limited by the imaging system, but instead by the thermal fluctuations of the atoms. We also demonstrate that electron ptychography can potentially reach a sub-nanometer depth resolution and locate embedded atomic dopants in all three dimensions with only a single projection measurement.

Transmission electron microscopy (TEM) plays critical roles in studying micro and nano-structures in many fields including physics, chemistry, structural biology, and material science. In the past two decades, progress in aberration-corrector optics for electron microscopes (*1, 2*) has significantly improved the quality of the imaging system, pushing the spatial resolution to sub-0.5 Å (*3, 4*). Nevertheless, these resolution limits in practical samples are only achievable in extremely favorable conditions. One major obstacle is that multiple electron scattering is unavoidable in samples thicker than a monolayer due to the strong Coulomb interaction between the beam electrons and the electrostatic potentials from the atoms (*5*). The multiple scattering changes the beam shape within the sample and leads to a complicated intensity distribution at the detector plane. Nonlinear or even non-monotonic contrast dependences on the sample thickness occur when imaging samples thicker than a few dozen atoms, which hampers direct structure determination by phase-contrast imaging. Dynamical electron scattering theories have been developed since H. Bethe (*6*) and quantitative structure-image interpretation usually relies on intensive image simulations and modeling (*7, 8*). Direct retrieval of the sample potential requires solving the nonlinear, inverse problem of multiple scattering. Efforts have been made through different approaches mostly based on Bloch wave theory by phasing different Bragg beams of crystalline samples (*9, 10*). Unfortunately, these approaches become extremely difficult for general samples with large unit-cells or aperiodic structures, because an incredibly large number of unknown structure factors needs to be determined. A more general approach has yet to be demonstrated.

Ptychography is another phase retrieval approach stemming back to Hoppe in the 1960's (*11*) and modern robust setups use multiple intensity measurements, usually a series of diffraction patterns collected by scanning a small probe across the extended sample (*12, 13*). No periodicity or symmetry constraints on the sample structure are required as a-priori knowledge. This approach has been widely used in visible light (*14*) and X-ray imaging communities (*15*). But until recently, electron ptychography has been severely limited by sample thickness and the limited detector performance in electron microscopy. Two-dimensional (2D) materials and the development of direct-electron detectors have led to a wider renewed interest (*16-20*). Electron ptychography for thin samples such as 2D materials has demonstrated an imaging resolution 2.5 times beyond the diffraction limit of the lenses, down to a 0.39 Å Abbe resolution recently (*18*). Such super resolution approaches can, however, only be applied reliably to samples thinner than a few nanometers, and the resolution is little different from conventional methods in thicker samples (*21, 22*). Such thin samples are practically difficult to achieve for many bulk materials, which currently limits applications to 2D-like systems, such as twisted bilayers (*20*). For samples thicker than the probe's depth of focus, multislice ptychography using multiple slices to represent the sample has been proposed (*23*). The structures of all slices can be retrieved separately. There are several successful demonstrations of multislice ptychography using either visible light (*24*) or X-ray (*25, 26*). Due to experimental challenges, However, only a few proof-of-principle multislice electron ptychography demonstrations (*27-29*) have been reported, limited either in resolution or stability. Here, we demonstrate multislice electron ptychographic reconstructions experimentally, recovering a linear phase response vs thickness and push the lateral resolution close to the intrinsic atomic size, limited by thermal fluctuations of the atoms themselves. We also show that a sub-nanometer depth resolution along the optical axis can potentially be achieved simultaneously with an ultimate lateral resolution.

The experimental setup is shown schematically in Fig. 1A. A focused electron probe is raster scanned across a slab-like sample, with one electron diffraction pattern recorded at each probe position using a high-dynamic range pixel array detector (fig. S1) (*30*). For thick samples, the probe function within the sample changes shape due to both wave propagation and strong dynamical scattering. A profile of the probe's evolution with depth into the sample is shown in Fig. 1B. In conventional ptychography, we approximate the wavefunction at the exit-surface of the sample as a multiplication of the incident wave function with a single projected sample function (*31*). For thick samples, however, portions of the sample at different depth positions are effectively illuminated with different wave functions due to the beam spreading by diffraction. Following the well-known Cowley-Moodie multislice solution of the electron dynamical scattering problem (*32*), the sample can be considered as many thin slices until each slice satisfies the multiplicative approximation (*31*). The whole scattering procedure is modelled as sequential scattering from each slice followed by a free-space progagation to the next slice. For the inverse problem, a similar multislice procedure can be adopted on each iteration as in multislice ptychography (*23*). The specimen potential for each slice is separately recovered through the phase of the transmission function, as illustrated in Fig. 1C (*30*).

We first compare the performance of multislice and single-slice electron ptychography on datasets simulated for crystalline $PrScO_3$, the same system used for our experimental measurements. For samples as thin as 8 nm, our simulation shows that multislice electron ptychography already gains a significant resolution improvement (Fig. 1D) over the single slice approximation. For thicker samples up to 30 nm, multislice electron ptychography demonstrates much stronger performance improvements, showing clearly the 0.59 Å separation of Pr-Pr dumbbells, while single-slice electron ptychography fails to even capture the basic structure. For even thicker samples such as 50 nm, reconstructions are still possible, but higher sampling densities (*26*) are required to obtain a convergent solution from multislice electron ptychography (fig. S2).

A key advantage of multislice ptychography is its linear phase dependence on the sample thickness. Multislice electron ptychography provides quantitative phase information with the phase increasing linearly as more layers are added into the sample. Figure 1E shows the phase change at different atomic positions from reconstructions with different sample thicknesses. Such linearity is crucial for retrieving three-dimensional structural information, especially for phase-contrast electron tomography which requires a monotonic contrast dependence of sample thickness (*33*). Nevertheless, conventional imaging methods such as scanning TEM (STEM) annular dark-field and annular bright-field or high-resolution TEM (HRTEM) images always have a nonlinear or even non-monotonic dependence on thickness (figs. S3 and S4), and extensive simulations are required for quantitative image interpretation. Additionally, fast data acquisition and low dose imaging capabilities outperforming conventional imaging techniques are retained in the multislice ptychography similar to single-slice ptychography (figs. S10 & S11) (*20*).

Now we turn to experimental data collected on a 300 keV Titan Themis with an electron microscope pixel array detector (EMPAD) (*34*). Full experimental details are given in the supplementary (*30*). Figure 2A shows one region of the phase image reconstructed using an experimental dataset from a $PrScO_3$ sample with a thickness of 21 nm projected along the [001] zone-axis, where the *c*-axis of $PrScO_3$ is taken as the longest axis; this is the non-standard *Pbnm* setting of space group #62. 42 slices were used during the reconstruction and each slice representing a thickness of 0.5 nm (*30*). The phase image clearly resolves all atoms in the structure and outperforms the state-of-the-art conventional electron microscopy methods (Fig. S6) in terms of contrast and resolution: all sublattices including both heavy-metal atoms, Pr and Sc, and light O atoms are resolved with a high contrast and signal-to-noise ratio. For this experimental data, both a partial coherence treatment of the electron probe (*20, 35*) and multiple slices (*23, 26*) are critical factors to obtain a high quality reconstruction (fig. S7).

The phase image of Fig. 2A is visually striking for its high spatial resolution, which is borne out by quantitative analysis. In real space, the Pr-Pr dumbbells with a separation of only 0.59 Å is resolved with a contrast of 63% (Fig. 2B), which is better than the contrast of two point objects separated at the Rayleigh criterion, 73%. Therefore, the image has a Rayleigh resolution much better than 0.59 Å. Nevertheless, as we will show later, the exact resolution can only be determined after considering the finite atomic size instead of assuming point objects (*36*). We can also resolve the O-Sc-O triple atom projections even though the light O atoms are only 0.63 Å (*37*) away from the heavier Sc atoms (Fig. 2C), and these cannot be resolved using conventional imaging techniques. The power spectrum from the Fourier transformation of the phase image (Figs. 2D and 2E) shows an isotropic information transfer that is larger than 4.39 Å$^{-1}$, corresponding to 0.23 Å in real space.

In order to estimate the spatial resolution of the ptychographic reconstruction, the intrinsic width of the atoms needs to be taken into account. The static projected potential of $PrScO_3$ (i.e., at zero Kelvin and neglecting zero-point thermal fluctuations) is very narrow (Fig. 3A). Nevertheless, the experiments were carried out at room temperature (300 K), and thermal fluctuations of atoms greatly broaden the potential (Fig. 3B). Additionally, imaging conditions, such as the finite illumination dose and maximum scattering angle of the collected diffraction patterns (*18*), will further impose a broadening factor in the reconstructed phase image (Fig. 3C). The combined effects of limited resolution and thermal fluctuations on the measured potential (Fig. 3D), each being roughly Gaussian in profile, can be added in quadrature and approximately modelled as a convolution of one Gaussian function with the static, frozen potential (*30*). The width of the Gaussian function is the total broadening factor of a ptychographic reconstruction, including the blurring of both the measurement technique and the thermal fluctuations of the sample. From the experimental data, the measured widths (full width at half maximum, FWHM) of each atomic column estimated from more than 60 atomic columns are 0.44 ± 0.01 Å, 0.45 ± 0.01 Å, and 0.54 ± 0.02 Å from Pr, Sc and O, respectively. By comparing the measured column widths with the Gaussian convolved potential, we can obtain the combined broadening factors (i.e., the FWHM of the convolved Gaussian) of 0.28 Å, 0.25 Å, and 0.34 Å for Pr, Sc, and O, respectively (fig. S12 and Table S1).

Thermal broadening factors of atoms can be calculated from Debye-Waller factors (DWFs) obtained from X-ray diffraction (XRD) measurements. We also measured DWFs from quantitative convergent beam electron diffraction (QCBED) (Table S2 and fig. S17) (*38*). The DWFs of Pr and Sc atoms from our QCBED measurements agree well with XRD measurements from $PrScO_3$ single crystals (*37*). For example, after converting to the FWHM of the thermal displacement, the thermal broadening factor for Pr from both XRD (*37*) and QCBED is 0.23 Å, compared to the measured total broadening of 0.28 Å for our reconstructed image. In other words, most of the measured broadening is already accounted for by the thermal vibrations in the sample. We estimate the residual instrumental contribution via quadratically subtraction, which gives the residual blurring (as a Gaussian FWHM) of our ptychographic reconstruction at Pr and Sc sites of 0.16 ± 0.01 Å and 0.15 ± 0.01 Å, respectively. We note that QCBED gives a larger DWF for oxygen site #2 (O#2) than does XRD. The resolution of O#2 is 0.23 ± 0.02 Å or 0.19 ± 0.02 Å, depending on whether the QCBED or XRD result is adopted. It is not surprising that the resolution estimated from different elements is different, because the quality of the ptychographic reconstruction at a finite illuminated dose is dependent on the scattering power of the object (*18*). Therefore, the Abbe resolution of our ptychographic reconstruction is better than 0.15 ± 0.01 Å and its Rayleigh criteria correspondence is 0.18 ± 0.01 Å (*30*). In all cases, the column width of the ptychographic reconstruction is mainly limited by the finite size of atoms determined by their thermal fluctuations instead of the imaging system itself.

Besides the resolution improvement, the precision for measuring the atomic positions is also significantly improved. Figure 3E shows the repeated measurements of Pr-Pr atomic distances. The standard deviation of the distance distribution is 0.7 pm, which indicates that we have achieved a sub-picometer precision simultaneously with the 16 pm resolution. More importantly, the positions of the light oxygen are also measured precisely. In contrast, these are not detectable by conventional annular dark field imaging, and their positions are not reliably recovered by annular bright field imaging (*39*). Since there are two different bond lengths in the distorted $ScO_6$ octahedra in $PrScO_3$, the Sc-O distance along two vertical directions, labelled as $d_1$ and $d_2$ on Fig. 3D, are not equal. The precision is close to 1 pm and the histogram in Fig. 3F shows distinguished $d_1$ and $d_2$ values, 2.030 ± 0.015 Å and 2.052 ± 0.013 Å, respectively. $d_1$ and $d_2$ only differ by ~ 2 pm and picometer precision is required to distinguish such a small difference. The high precision from conventional imaging techniques, however, can only be realized for heavy metals, and multiple frames acquisition together with sophisticated image registration algorithms (*40*). The high precision together with the high resolution measurement for both heavy and light atoms are crucial for correlating structures and functionals in materials. It is important to stress that multislice ptychography does not rely on lateral periodicity, so it can be applied equally to defects and grain boundaries as it can to single crystals.

Multislice electron ptychography also allows for three-dimensional structure determination since it iteratively retrieves the sample structure at different layers. First, we check the reconstructed structures at different slices from the experimental results; three example slices are shown in Fig. 4A (all slices given in Movie S1). We find that the slices at the beginning and at the end (Movie S1) show very small phase shifts. In middle slices, the phase images show strong contrast and clear structural features. Figure 4B shows a depth profile of the phase change cut along the Pr-O direction (marked as a dashed line on Fig. 4A). This depth evolution comes from the fact that the sample is close to a parallel-sided film and there are vacuum layers above and beneath the film (Fig. 4C). The electron beam changes shape as a function of depth into the sample, which is properly accounted for in the multislice electron ptychography algorithm, and as a result recovers a phase shift that is linearly proportional to the electrostatic potential of the sample at each different depth slice. The broadening of the depth profile from each surface of the rectangular slab is fitted by an error function and gives a depth resolution of ~ 3.9 nm estimated from the Pr sites (Fig. 4D) (*30*). This is better than the aperture-limited depth resolution of 5.1 nm from conventional optical sectioning imaging (*30*).

We also find through simulations that multislice electron ptychography can enable the detection and locating of single atomic dopants in all three dimensions. We constructed a structural model by introducing single dopants in a 15 nm-thick $PrScO_3$ crystalline matrix and generated diffraction data using mutislice simulations (*30*). Dopants of different elements are put at different interstital or substitutional sites and different depth positions. Figure 4E shows the reconstructed phase images from three slices out of 30 total slices (Movie S2) from the two Pr dopants at depths differing by 3 nm. The depth profile across the dopant (Fig. 4F) shows well localized contrast from the two single dopants in all three dimensions. The depth resolution estimated from the full width at 80% of the maximum (FW80M) at the dopant peak in the depth profile (Pr2 in Fig. 4G) is 0.9 nm. Depth profiles from single Pr, Sc, and O dopant (Fig. 4G) are similar and show FW80M in depth of 0.9 nm, 1.6 nm, and 2.0 nm, respectively. This also shows that the depth resolution depends on the scattering power of the dopant, which in turn depends on the atomic number, as noted for the lateral resolution discussed earlier. A strong contrast and good depth resolution are also retained for single dopants on the atomic columns (fig. S14). Importantly, the depth resolution from multislice electron ptychography exceeds the aperture-limited resolution (*41*), and it has an illumination-dose dependence (Fig. 4H). Additionally, the depth resolution depends on contrast and transverse resolution and can be potentially improved by collecting electrons scattered to higher angles and using a more converging or diverging electron beam (*26*).

Since conventional imaging methods are based on projections through the entire sample, single dopants embedded in a relatively thick matrix usually show little or no contrast (fig. S16). Only dopants with large atomic numbers in a light element matrix and a very thin sample may be detectable. For example, a Sb atom in a silicon matrix can only be distinguished using STEM annular dark-field images in samples thinner than 5 nm (*42*). Furthermore, there is strong multiple electron scattering in crystalline samples and the resultant rechanneling of the electron beam may result in incorrect column locations for dopants via conventional optical sectioning methods (*43*). These artifacts also have consequences for electron tomography, the most commonly used technique for determining three-dimensional structures of samples (*44*), because poor or incorrect contrast from single dopants in the projection images used for electron tomography hinder their identification and correct localization. The recovered linear signal provided by our full inversion of electron multiple scattering via multislice electron ptychography addresses these deficiencies. Therefore, multislice electron ptychography provides a powerful tool for locating single dopants at more than double the resolution in all three dimensions compared to conventional approaches. Combined with tomography, robust atomic-resolution details of defect clusters should be possible in all three dimensions.

**Acknowledgments: Funding:** Research was supported by the National Science Foundation under grant DMR-1539918 (PARADIM Materials Innovation Platform in-house program). Electron microscopy performed at Cornell Center for Materials Research facility supported by National Science Foundation under grant DMR-1719875. **Author contributions:** Z.C. and D.A.M. initiated and conceived the research; Z.C. performed the experiments and data analyses under the supervision of D.A.M.; Z.C. and Y.J. implemented the mixed-state multislice ptychography algorithms based on the initial ptychography codes from M.O. and M.G.-S. Y.-T.S. performed the Debye-Waller factors measurements. M.E.H. prepared the sample under the supervision of D.G.S. who also provided advice on XRD structural refinements. I.H. grew the PrScO$_3$ single crystal under the supervision of S.G. by the Czochralski method. Z.C. wrote the manuscript with revisions from D.A.M. and inputs from all authors. **Competing interests:** Cornell University has licensed the EMPAD hardware to Thermo Scientific. **Data and materials availability:** Raw experimental data of the results presented in the main text and supplementary materials is available after the publication from PARADIM, a National Science Foundation Materials Innovation Platform [https://doi.org/xxxxxx]. The codes developed at Cornell University are published on GitHub, muller-group-cornell [https://github.com/muller-group-cornell]. The ptychography reconstruction toolkit, PtychoShelves developed at Paul Scherrer Institut, Switzerland, is available on the website [https://www.psi.ch/en/sls/csaxs/software].

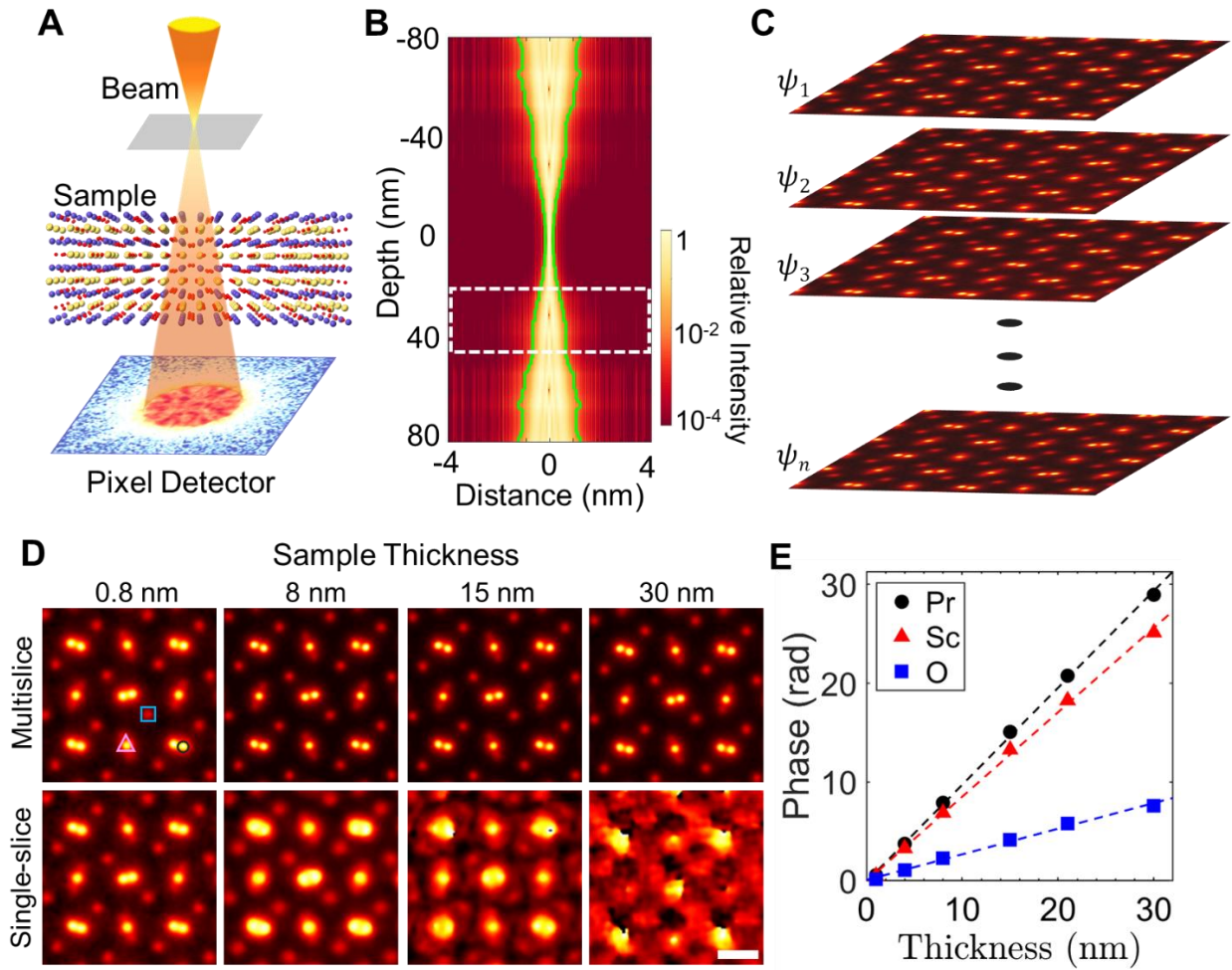

**Fig. 1. Principle of multislice electron ptychography.**

(**A**) Experimental setup for electron ptychography. (**B**) Intensity distribution of electron probe during free-space propagation. Green lines mark the boundaries of the probe intensity at one percent of the maximum. The sample position is illustrated by a dashed rectangle. (**C**) Transmission functions of multiple objects at different depth positions reconstructed via multislice electron ptychography, $\psi_i$ ($i$=1, 2, …, $n$). (**D**) Total phase images summed from each slice reconstructed via multislice electron ptychography using simulated detector data for PrScO$_3$ [001] samples with different thicknesses. The slice thickness used is 0.5 nm. Corresponding phase images reconstructed via single-slice electron ptychography are also shown for comparison. The scale bar is 2 Å. (**E**) Phases at Pr, Sc, and O atomic sites (marked by a circle, triangle, and rectangle on the first image of (D), respectively) from samples with different thicknesses obtained via multislice ptychographic reconstructions. Linear fits are shown as dashed lines.

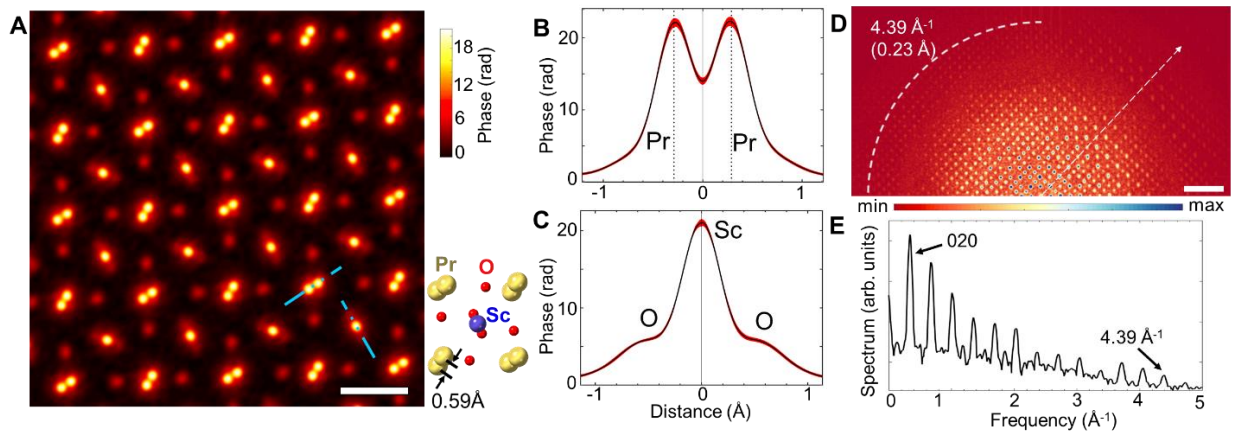

**Fig. 2. Multislice electron ptychographic reconstruction of PrScO$_3$.**

(**A**) Total phase image, from summing all slices, of [001]-oriented PrScO$_3$ from a multislice electron ptychographic reconstruction of experimental data. A structural model of PrScO$_3$ is shown as an inset. The scale bar is 2 Å. (**B**) and (**C**) Phase profiles across Pr-Pr dumbbells and O-Sc-O directions, respectively. The red shading illustrates the variation from about 60 profiles at different atomic columns and black lines are their average. Example positions are marked as dashed blue lines on (A). (**D**) Intensity of Fourier transformation of the reconstructed phase image. The scale bar is 1 Å$^{-1}$. (**E**) Power spectrum along the direction marked by the dashed arrow in (D). The intensity of power spectrum in (D) and (E) is weighted as a power of 0.2 to highlight the high-frequency information transfer. The Bragg peak of 020 is labelled on (E).

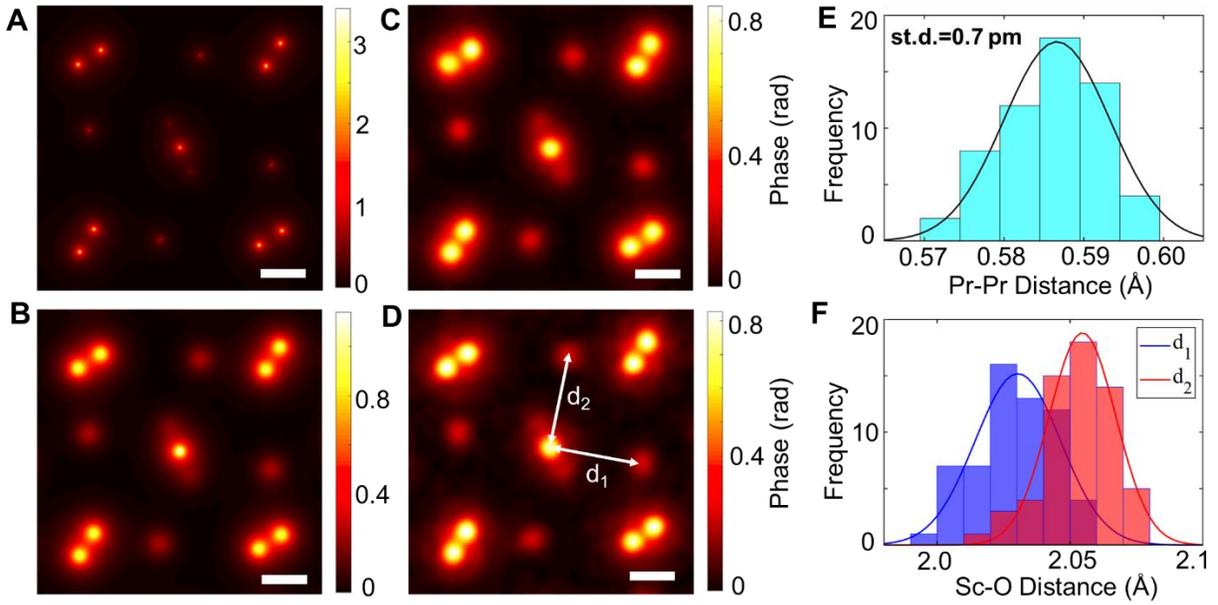

**Fig. 3. Spatial resolution and measurement precision from multislice electron ptychography.**

(**A**) Projected electrostatic potential of PrScO$_3$ in the absence of thermal vibrations. (**B**) Projected potential after the thermal broadening at 300 K. (**C**) Ptychographic reconstructed potential from simulated data including thermal fluctuations from 21 nm thick PrScO$_3$ sample. (**D**) A cropped region of the ptychographic reconstructed potential from experimental data. The potential in both (C) and (D) is normalized to the unit-cell thickness (8.03 Å). (**E**) Histogram of projected Pr-Pr distances from experimental measurements. The average Pr-Pr distance is 0.586 ± 0.007 Å. (**F**) Histogram of the projected Sc-O bond-length from experimental measurements. $d_1$ and $d_2$ indicate different Sc-O distances along the two vertical directions as labelled on (D). The average Sc-O distances are 2.030 ± 0.015 Å and 2.052 ± 0.013 Å, respectively. The scale bar in (A)-(D) is 1 Å.

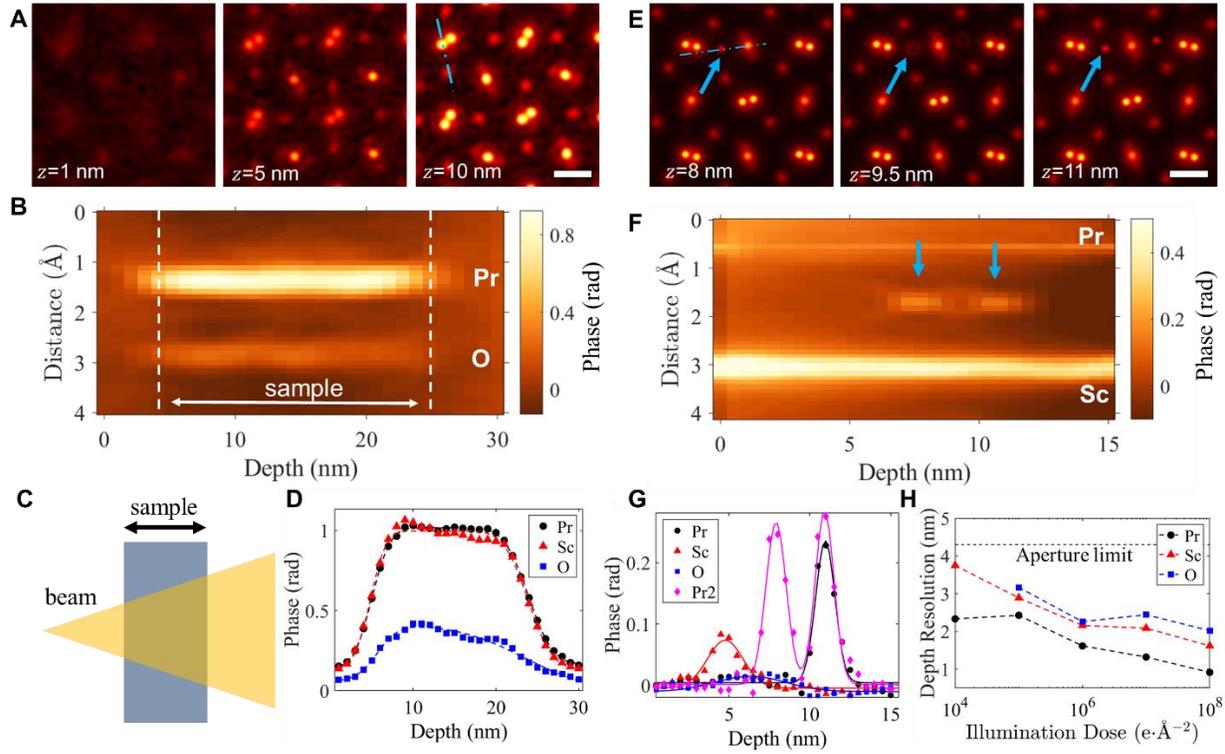

**Fig. 4. Depth sectioning of multislice electron ptychography.**

(**A**) Ptychographic reconstructed phase of three selected slices at different depths, $z$, from experimental measurements of $PrScO_3$. The sample thickness is initially assumed as 30 nm during the reconstruction, but the retrieved sample thickness is 21 nm. (**B**) Depth profile of phase intensity along the Pr-O direction marked on (A) by the blue dash-dotted line. (**C**) A cartoon showing the electron beam spreading into the sample. (**D**) Measured depth profile of the phase at different atomic columns in (A). The full width at 80% of the maximum from the error function fittings are 3.9 nm, 3.5 nm, and 4.2 nm of Pr, Sc, and O sites, respectively. (**E**) Simulated ptychographic reconstructions of the phase for three selected slices from a model with two interstitial Pr atom dopants located at a depth separation of 3 nm. (**F**) Depth variation of phase intensity across the Pr dopants. The two Pr dopants are marked with arrows. (**G**) Depth variation of the phase for different dopant positions. The solid lines are Gaussian fits. The results in (E) to (G) are for a modeled illuminated dose of $10^8$ e·Å$^{-2}$. (**H**) Computed depth resolution for multislice electron ptychography as a function of illuminated dose. The aperture-limited depth resolution (4.3 nm) is shown as a dashed horizontal line. The scale bar in (**A**) and (**E**) is 2 Å.

# Supplementary Materials for

Electron ptychography achieves atomic-resolution limits set by lattice vibrations

Zhen Chen*, Yi Jiang, Yu-Tsun Shao, Megan E. Holtz, Michal Odstrčil, Manuel Guizar-Sicairos, Isabelle Hanke, Steffen Ganschow, Darrell G. Schlom, David A. Muller*

Correspondence to: zhen.chen@cornell.edu (Z.C.); david.a.muller@cornell.edu (D.A.M.)

**This PDF file includes:**

    Materials and Methods
    Figs. S1 to S17
    Tables S1 to S2
    Captions for Movies S1 to S2
    References

**Other Supplementary Materials for this manuscript include the following:**

    Movie S1 and S2 (.mp4)

# Materials and Methods

Experimental conditions

Transmission electron microscopy (TEM) specimen was prepared from a PrScO$_3$ single crystal with Ga+ ion beam using the standard lift-out method on a focus-ion-beam (FIB, FEI Strata 400). Subsequent lowering the energy of the ion beam down to 2 keV was used to reduce sample damage during thinning. The thinnest part of the wedge-shaped sample is smaller than 10 nm.

TEM experiments were performed on an aberration-corrected electron microscope (Thermo Fisher Scientific, Titan Themis) with a probe-forming aperture semi-angle of 21.4 mrad at 300 keV. The four-dimensional scanning transmission electron microscopy (4D-STEM) dataset from PrScO$_3$ single crystal along [001] zone-axis was acquired using a high-dynamic range electron microscope pixel array detector (EMPAD) with 128 × 128 pixels (*1*). We adopted a defocused electron probe focusing ~ 20 nm above the sample top-surface plane. The acquisition conditions of 4D dataset are as follows: the scan step size is 0.41 Å, total scanning points are 128 × 128 with a scanning area of 5.2 × 5.2 nm$^2$, the beam current is ~ 33 pA, the acquisition time per diffraction pattern is 1 ms with an additional 0.86 ms readout time, giving an illumination dose of 1.2 × 10$^6$ e·Å$^{-2}$. A proper camera length was chosen to ensure the maximum scattering angle of the acquired diffraction is about 2.5 times of the probe-forming semi-angle with a sampling of 0.823 mrad per pixel in the diffraction patterns.

Reconstruction algorithms

Modern electron ptychography usually uses the diffraction patterns acquired from partly overlapped positions in the scanning setup (Fig. S1). For thin samples such as 2D materials, electron ptychography is based on the approximation that the wavefunction at the exit-surface plane of the sample, $\psi_j(\mathbf{r})$, is a multiplication of the incident wavefunction and a single complex transmission function (*2*),

$$\psi_j(\mathbf{r}) = P(\mathbf{r} - \mathbf{r}_j)O(\mathbf{r})$$

where $P$ is the probe function, $O(\mathbf{r}) = A(\mathbf{r})\exp(i\sigma V(\mathbf{r}))$ is the transmission function or the object function of the sample, $\mathbf{r}$ is the real-space coordinates and $\mathbf{r}_j$ is the probe position of the *j*-th diffraction pattern. The phase of $O(\mathbf{r})$ is the sample electrostatic potential, $V(\mathbf{r})$, containing the sample structure information, besides an interaction constant $\sigma$, only depending on the energy of the incident electron beam. $A(\mathbf{r})$ is the amplitude showing the absorption of the electrons in the sample. Atomic resolution electron ptychography measurements on 2D materials that outperform conventional imaging modes have been demonstrated recently (*3-5*).

However, for samples thicker than a few nanometers where significant multiple scattering occurs, the multiplicative approximation no longer holds. The thickness limit, $T$, is empirically given as $T < \frac{1.3\lambda}{\theta_{\max}^2}$ if only the probe propagation effects are considered (*6*), where $\theta_{\max}$ is the maximum scattering angle of the diffraction pattern and $\lambda$ is the wavelength of electron beam. Typical imaging conditions in TEM gives a thickness limit of only a few nanometers (*5*). Multiple scattering in conventional imaging techniques has been intensively investigated via numerical simulations (*7-9*). Following the well-known multislice theory in TEM (*10*), schematically shown in Fig. S1D, the object function of the sample can be divided into multiple slices. For each slice, there is a separate potential $V_i(\mathbf{r})$ ($i$=1,2, ..., $n$, $n$ is the number of slices) representing the projected potential of the slice within slice thickness of $\Delta z_i$. If the slice is sufficiently thin, the multiplicative approximation holds. Based on this idea, multislice ptychography, firstly proposed in 2012 (*11*) inversely retrieves the object function at each slice and at the same time the incident probe function.

The total object function can be written as,

$$O(\mathbf{r}) = \prod_{i=1}^{n} O_i(\mathbf{r}) = (\prod_{i=1}^{n} A_i(\mathbf{r})) \exp(i\sigma \sum_{i=1}^{n} V_i(\mathbf{r})),$$

where $\prod$ denotes multiplications and $\sum$ denotes summations. Therefore, the total potential or phase is the summation of phase from every slice. We implemented multislice electron ptychography in the framework of the generalized maximum-likelihood method with gradient descent optimization algorithms (*5, 12, 13*). A Fresnel operator for the wave propagation within the sample was used. The partial coherence of the electron probe is treated via a mixed quantum state (*14*) similar to that used for conventional electron ptychography for thin samples (*5*). Each probe mode propagates separately through all slices to form separate exit-surface wavefunction, and the total diffraction pattern is the incoherently summation of the diffraction patterns from each probe mode.

Figure S5 shows a multislice electron ptychography reconstruction from a 128 × 128 experimental diffraction patterns, including the phase and amplitude of the sample as well as the intensity of all 12 probe modes. We find that for the experimental data from $PrScO_3$ sample, both multiple slices of the object and a mixed-state probe must be employed in order to obtain a high-resolution reconstruction (Fig. S7). Similar to the single slice ptychography demonstrated recently (*5*), probe position refinement and variation of probe wavefront is also critically important for multislice ptychography. The difference here is that there are many slices. In principle, the probe position can be refined from the gradient of the object or probe in any slice (*6, 12*). But we find that position refinement based on the object function of some middle slice works the best.

The number of slices, *n*, depends on the total sample thickness and the slice thickness (or more accurately, the propagation distance of each slice). We tested the effects of the slice thickness on the quality of the reconstruction with a fixed total thickness of 20 nm using simulation data from 20-nm-thick $PrScO_3$ (Fig. S8). We find that a high-resolution reconstruction can be obtained when slice thickness, $\Delta z_i$ is smaller than ~ 2 nm. Further resolution improvements of the total object function can be seen when the slice thickness is reduced, down to 0.5 nm. However, smaller slice thickness means that a larger number of slices is required for a fixed sample thickness and more object functions need to be reconstructed. Therefore, we chose $\Delta z_i$ =0.5 nm for all the reconstructions except for those explicitly specified. It is worth mentioning that the best slice thickness may be dependent on the convergence angle of the probe (*6*) and the scattering power of the sample. We note that for thick samples or experimental data with large errors from the probe position measurements, artifacts may appear if too many slices are used. This is mainly due to the intrinsic ambiguity to decouple the object and the probe when depth of focus of probe is much larger than the slice thickness (*11*). To avoid such artifacts, we implemented a new regularization algorithm. For each iteration, low spatial frequencies information of object functions from different slices are weighted by a regularization factor in Fourier space after unwrapping the phase of object layers (*13*). This is basically to model the information mixing between slices due to small phase factors of probe at low spatial frequencies in Fresnel propagator (*6*). A strong regularization making all slices near identical helps for probe position refinement and can be employed as a first step to get a better initial start. Additional reconstructions with lighter regularization parameters allow for a good sectioning of structural features at different depths, such as identifying the single atomic dopant.

We treat the sample thickness as a free parameter. The correct thickness can be determined from two different criteria (Fig. S9). First, the numerical difference between the diffraction patterns generated from ptychography and the measured ones reaches a minimum when the thickness is correct because a best match to the reconstruction is achieved at this condition. Second, the total phase from all slices reaches about the maximum at the sample thickness. Maximum phase negligibly increases when the sample thickness further increases because the depth part of the sample larger than the real thickness is basically vacuum layers and has negligible contribution to the total phase. The sample thickness determined from these two separate strategies agrees very well and both give the correct thickness when tested by simulation (Fig. S9).

For thick samples, it may be difficult to get a converged reconstruction from multislice electron ptychography. Firstly, object functions at many slices must be determined, and more unknown parameters may lead to insufficient data diversity required for the phase retrieval problem. Secondly, the electron intensity distribution within the sample can be very diffused due to the multiple scattering and beam propagation. Under the conditions studied in this article, it is relatively straightforward to obtain high-resolution reconstructions from samples thinner than 30 nm in $PrScO_3$. With increasing sample thickness, the electron probe spreads laterally to larger sample regions due to the longer propagation distance within the sample, and thus each diffraction pattern includes scattering events further away from the probe position. Consequently, larger scanning regions in real space and better samplings in diffraction space are needed for thicker samples, such as the simulated 50 nm thickness shown in Fig. S2.

Multislice electron ptychography also retains the advantage of faster data acquisition as with single-slice out-of-focus electron ptychography (*5*) compared with conventional scanned imaging techniques. As we know, the real space sampling in the ptychographic reconstruction is determined by the largest scattering angle of the diffraction patterns instead of the scan step size, *i.e.*, the interval between two neighboring positions where the diffraction pattern is acquired. Scan step sizes as large as 4 Å using EMPAD detectors can be used in conventional single-slice electron ptychography for thin samples such as 2D materials (*5*). For our case, the real-space resolution of the reconstruction is about 1/10 of the scan step and 1/20 of the sampling assuming Nyquist sampling, and a factor of four hundred times can potentially be increased in the data acquisition speed. For thick samples, there are more features in the diffraction patterns due to strong multiple scattering. Furthermore, more unknowns about the multiple object functions also require a higher data diversity. Therefore, a higher sampling for both diffraction patterns and scanning plane is expected. Whether high speed data acquisition retains needs to be revisited. Using the experimental data from a 21-nm-thick PrScO$_3$ sample, we find that high-quality reconstructions can still be achieved with a scan step size up to ~ 1 Å, which is ~ 5 times of the resolution or , ~ 10 times larger than the pixel size in the reconstruction (0.093 Å) (Fig. S10). This gives about one hundred times gain in data acquisition speed, which is smaller than that for thin samples but still significant. It is possible that an even larger scan step size can be used if the probe coherence is improved, number of camera pixels is increased, or the local fluctuation of the probe position during data acquisition is reduced.

Another advantage of electron ptychography is its dose efficiency. In conventional single-slice electron ptychography, low dose imaging has been demonstrated using 2D materials or biological materials (*4, 5, 15, 16*). For thick sample, phase contrast imaging techniques such as conventional high-resolution TEM (HRTEM) (*7*) or in scanning TEM (STEM) such as integrated center-of-mass/differential phase contrast (iCoM/iDPC) (*9, 17*) or single-slice electron ptychography (*18, 19*) do not show a good resolution or direct interpretable structural information. We usually use high angle annular dark-field (HAADF) or annular bright field (ABF) images to resolve heavy or light elements, respectively. We compare the dose efficiency between HAADF, ABF and multislice electron ptychography via simulations for a 15 nm thick PrScO$_3$ sample (Fig. S11). Ptychography achieves a factor of more than ten times dose reduction for identifying light atoms such as oxygen atoms and 3-5 times for heavy atoms such as Pr atoms.

The main limitation for routine applications of multislice electron ptychography is the computational speed of the reconstruction. The required computational resources are mainly dependent on the sample thickness, number of probe modes and number of pixels in each diffraction pattern. For the reconstruction in Fig. S5 using 256 × 256 pixels of the detector, 128 × 128 diffraction patterns, 12 probe modes and 42 slices, ~ 45 minutes per iteration are needed on a Nvidia Quadro P5000 GPU card. We only need fewer than 100 iterations to get a converged solution if the initial probe is firstly retrieved from a small sub-region of the same dataset. Further optimization of the codes, especially better parallelization, is needed to realize live reconstructions during data acquisition.

Resolution and thermal fluctuations

Resolution is usually defined as the width of the point spread function (PSF) imposed by the imaging system (*20*). The image, $I(\mathbf{r})$ is the convolution of the object by the PSF,

$$I(\mathbf{r}) = PSF \otimes O(\mathbf{r}).$$

For a point object, expressed as a Dirac delta function, $\delta(\mathbf{r})$, the width of the PSF is the width of the final image, since the final image, $I(\mathbf{r})$ equals to the PSF. However, the atomic potential is not point object but has a finite width. For the potential from the ptychographic reconstruction, $V(\mathbf{r})$,

$$V(\mathbf{r}) = PSF \otimes V_T(\mathbf{r}),$$

where $V_T(\mathbf{r})$ is the potential at room temperature of $T$=300 K. The PSF can be approximately considered as a Gaussian function,

$$PSF = a_{pty} \exp\left(-\frac{r^2}{2\sigma_{pty}^2}\right).$$

The full-width at half-maximum (FWHM) of the PSF can be defined as the resolution, approximately following the Abbe resolution,

$$d_{pty} = 2\sqrt{2\ln(2)}\,\sigma_{pty}.$$

The thermal broadened potential at 300 K can be calculated from the Debye-Waller factors (DWFs), $B$ and the potential at 0 K, $V_0(\mathbf{r})$ (20),

$$V_T(\mathbf{r}) = a_1 \exp\left(-\frac{r^2}{2\sigma_{DWF}^2}\right) \otimes V_0(\mathbf{r}),$$

where $\sigma_{DWF} = \sqrt{\frac{B}{8\pi}}$. The width of the thermal broadening, $d_{DWF} = 2\sqrt{2\ln(2)}\,\sigma_{DWF}$. Therefore, $V(\mathbf{r}) = G(\mathbf{r}) \otimes V_0(\mathbf{r})$, where, the total broadening can be expressed as a Gaussian function, $G(\mathbf{r}) = a\exp\left(-\frac{r^2}{2\sigma_{tot}^2}\right)$, and $\sigma_{tot} = \sqrt{\sigma_{pty}^2 + \sigma_{DWF}^2}$, the total width $d_{tot} = 2\sqrt{2\ln(2)}\,\sigma_{tot}$. Therefore, the width of the PSF can be determined via $d_{pty} = \sqrt{d_{tot}^2 - d_{DWF}^2}$.

To determine the resolution of the ptychographic reconstruction from the experimental data, we firstly measured the width of each atomic column from the reconstructed phase image. We used Voigt functions to fit the profiles across the atomic positions. The width (FWHM) of Pr, Sc, and O (O#2) are 0.44 ± 0.01 Å, 0.45 ± 0.01 Å and 0.54 ± 0.02 Å, respectively, statistically averaged from more than 60 Pr-Pr dumbbells, Sc or O columns. Then we calculated the potential at 0 K using the scattering factors from relativistic Hatree-Fock atomic wavefunctions (21, 22). Starting from the simulated potential at 0 K, we can obtain a broadened potential by convolving Gaussian functions of different widths. The width of the atomic columns in the broadened potential is approximately linearly dependent on the width (FWHM) of the Gaussian function (Fig. S12D). A total broadening factor ($d_{tot}$) can be determined by matching the width of the broadened potential with the experimental ones. From the Debye-Waller factors refined from X-ray diffraction of single crystals in (23), the thermal broadening widths ($d_{DWF}$) from Pr, Sc, and O columns are 0.23 Å, 0.20 and 0.28 Å, respectively. The resolution ($d_{pty}$) of our ptychographic reconstruction estimated from Pr, Sc, and O columns are 0.16 Å, 0.15 and 0.23 Å, respectively. Such a resolution estimation from the width of the point spread function is close to the Abbe resolution definition. Therefore, the corresponding Rayleigh resolution are 0.20 Å, 0.18 Å, and 0.28 Å. All related quantities are listed in Table S1.

We also measured Debye-Waller factors (DWFs) of $PrScO_3$ using quantitative convergent electron diffraction (QCBED). Firstly, the electron beam energy of the microscope that is crucial for DWFs measurements was calibrated using energy filtered (EF) CBED pattern from a standard silicon sample near the [320] zone axis (24). EF-CBED was performed using a post-column Gatan GIF Tridiem energy filter through an energy window of 10 eV, with a probe size of ~1 nm at FWHM (full-width at half-maximum) on a Titan Themis electron microscope. The beam energy was determined as 300.4 keV by comparing the intersection of experimental high-order Laue zone (HOLZ) lines with calculated HOLZ lines (*25*). The DWFs are estimated by matching the experimental EF-CBED patterns with Bloch-Wave simulations using the *EXTAL* program (*24, 26, 27*). To improve the precision of DWFs, we deliberately tilted the sample to a systematic row orientation to excite the HOLZ Bragg reflections. The CBED and the fitting results are shown in Fig. S17. The DWFs from our QCBED are listed in Table S2. Compared with previous X-ray diffraction (XRD) measurements, we find that our measurements are similar to that from XRD of single crystals in 2009 (XRD2009) (*23*) but have some deviations from that of XRD measurements from polycrystals in 2004 (XRD2004) (*28*). Compared to XRD2009, the only obvious difference is the DWF of oxygen site #2, which is not surprising since XRD has large uncertainties in determining DWFs from light elements. Therefore, the Abbe resolutions of our ptychographic reconstruction estimated using DWFs from QCBED are 0.16 Å, 0.15 and 0.19 Å from Pr, Sc, and O columns, respectively, and the corresponding Rayleigh resolutions are 0.20 Å, 0.18 Å, and 0.23 Å.

Electron diffraction simulations

To investigate the performance of multislice electron ptychography in different experimental conditions, such as the sample thickness or illumination dose, we simulated 4D-STEM diffraction patterns using $PrScO_3$ sample as a model system and multislice simulation program, *μSTEM* (*29*). Projected potential was constructed from $8 \times 8$ unit-cells with an area of $4.49 \times 4.62$ nm$^2$ under a periodic boundary condition. Thermal diffuse scattering is included under the frozen-phonon approximation with 20 phonon passes. Debye-Waller factors of $PrScO_3$ from X-ray diffraction refinement results (*23*) were adopted for simulations of crystalline samples. A larger Debye-Waller factor of Pr ($B$=0.12 Å$^2$) in dopants simulations for depth resolution studies is used to model a deeper potential. Parameters such as beam energy and probe forming semi-angle from the experiments were used in the simulations. A series of different sample thickness was performed. For sample thickness from 0.8 nm to 30 nm, we used a probe focusing 20 nm above the sample, and generated $35 \times 36$ diffraction patterns with $256 \times 256$ total pixels per pattern and a sampling of 0.856 mrad per pixel during the ptychographic reconstructions. But for the sample thickness of 50 nm in Fig. S2, a better sampling with 0.428 mrad per pixel and $512 \times 512$ total pixels for a real-space sampling of ~ 0.09 Å is required. Furthermore, we find that more diffraction patterns from a larger scanning area ($68 \times 70$ diffraction patterns) are required to obtain a good convergence during ptychographic reconstruction for the 50 nm thick sample, because the electron probe spreads to larger sample regions due to longer propagation distance within the sample. We also used a smaller defocus value of 10 nm to ensure that the beam is not too diffused within the sample. Scan step for all simulations is about 0.46 Å and the pixel size in the final reconstructions is 0.090 Å. Poisson noise was added to model the finite illumination dose conditions. Illumination dose used for Fig. 1 and 2 is $10^6$ e·Å$^2$, which is close to experimental data ($1.2 \times 10^6$ e·Å$^{-2}$).

Depth resolution of multislice electron ptychography

The spatial resolution along the optical axis, or depth resolution, $dz$, is limited by the probe forming aperture, $\alpha$, in conventional imaging techniques. Using the full width at 80% of maximum

(FW80M) intensity of the probe along the optical axis, the depth resolution of incoherent imaging modes is $dz = \frac{\lambda}{\alpha^2}$ (30). For the 300 keV electron beam and $\alpha = 21.4$ mrad used in this work, the aperture limited depth resolution is 4.3 nm. Chromatic aberration of the microscope also introduces a defocus blurring, $df = C_c \frac{\Delta E}{E_0}$, where $C_c$ is the chromatic aberration coefficient, $\Delta E$ is the energy spread of the electron beam with total energy of $E_0$. Here in our electron microscope, $C_c = 2$ mm, $\Delta E = 0.4$ eV (FW80M), and $E_0 = 300$ keV, and then $df = 2.7$ nm. These effects should be convolved with each other. Therefore, the total depth resolution from conventional imaging techniques is approximated by adding these two terms in quadrature to give a combined depth resolution of 5.1 nm.

We investigated the possible depth resolution at the atomic resolution regime in multislice electron ptychography using two different methods. First, we examined the depth broadening of the phase change at the sample top and bottom edges, as the sample can be considered as a parallel slab shape. Depth resolution can be measured as the width of the step function, which can be fitted using an error function at each side of the sample. We adapted the width of FW80M and the averaged value from both surfaces using the phase profiles at Pr atomic column positions. The results from experimental data have been discussed in the main text Fig. 4. Depth resolution from simulated data for a 20-nm-thick PrScO$_3$ sample is 2.0 nm using a dose of $10^6$ e·Å$^2$ similar to the experiment (Fig. S13). Compared to the results from the experiments, the depth resolution from simulation is about two times better. There are two main reasons. Firstly, there is further depth blurring due to the chromatic aberrations (31) and partial spatial coherence in the experimental data. Secondly, a stronger regularization imposing a furuther broadening between slices has to be employed to get a good reconstruction for the experimental data. Because uncertainties from probe partial coherence and refined probe positions introduce additional difficulties during decoupling the object and probe.

We also studied the depth resolution by directly measuring the depth positions of single dopants by model simulations. We introduced single dopants into a 15 nm thick crystalline matrix of PrScO$_3$. There are two different types of dopants. Firstly, single Pr, Sc, and O atoms are put in the intersitial sites between atomic columns of the matrix, ~ 1 Å apart from the nearest atomic columns in the [001]-axis projection. The depth positions ($z$) of Pr, Sc, and O single dopants are at 10.7 nm, 4.6 nm, and 7.5 nm, respectively. We also put two Pr atoms at the same lateral position but 3 nm apart along the optical axis to directly demonstrate the real-space depth resolution. Two Pr atoms are at $z$=7.5 nm and $z$=10.5 nm, respectively. We can resolve the two Pr dopants in real space if the illumination dose is higher than $10^6$ e·Å$^2$ (Fig. S15). Secondly, we replace one Pr atomic site ($z$=7.4 nm) with one Sc atom and one Sc atomic site ($z$=7.5 nm) with one Pr atom to demonstrate whether the dopant in the atomic column can be resolved. We can also distinguish the substitutioinal dopants in high illumination dose condition (Fig. S14C and D). However, there are extradinary difficulties for conventional optical sectioning in such a condition (32). The phase images of all slices are shown in Movie S2, sequentially from top to bottom of the sample. All the single dopants are not visible in conventional STEM images, such as HAADF and ABF images (Fig. S16), even with a very high signal-to-noise in high illumination dose conditions.

Supplementary Figures:

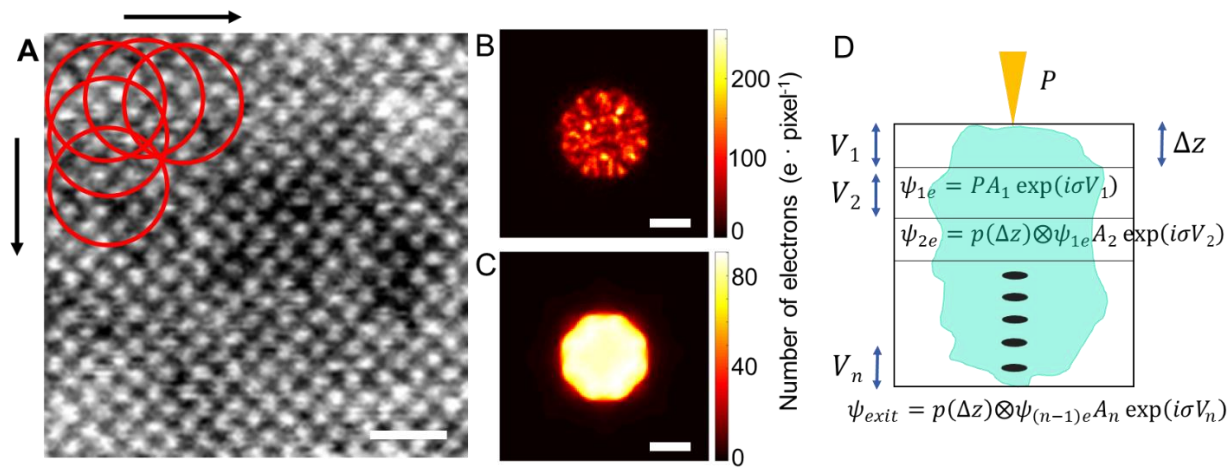

**Fig. S1.**

Workflow of multislice electron ptychography. (A) Schematic of the scanning data acquisition. The defocused probe, illustrated as red circles, scans across the sample. An annular dark-field image synthesized from the scanning diffraction patterns is shown. (B) One diffraction pattern from a single probe position. (C) Position averaged diffraction pattern from the whole region. (D) Representative illustration of multi-slice electron scattering theory. $V_i$ is the electrostatic potential of each slice, $i$, $\sigma$ is the interaction constant, $\Delta z$ is the thickness of each slice. $P$ is the wave function of incident probe, and $\psi_{ie}$ ($i$=1, 2, …, $n$) is the exit-surface wave function of slice $i$. The scale bar in (A) is 1 nm and in (B) and (C) is 1 Å$^{-1}$.

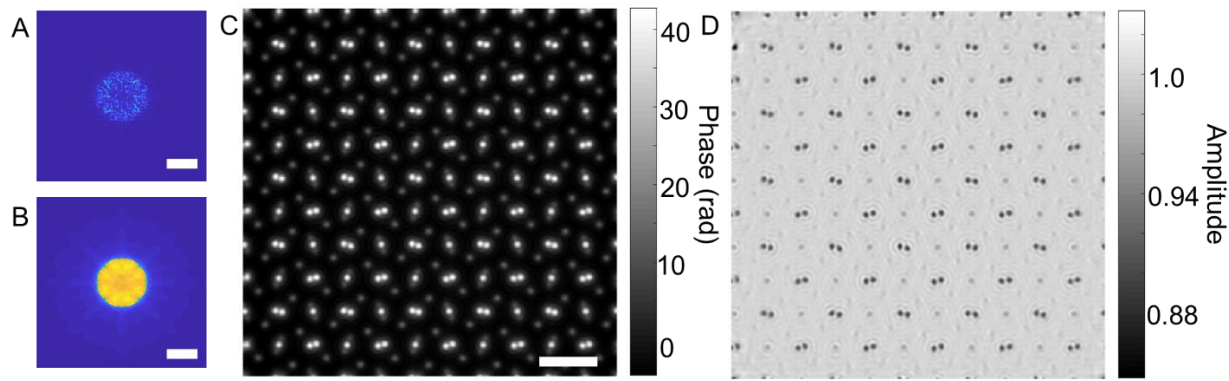

**Fig. S2.**

Multislice ptychographic reconstruction of a 50 nm thick PrScO$_3$ sample from simulated data with a probe focusing 10 nm above the sample. The sampling in the diffraction pattern is 0.428 mrad per pixel, which is two times smaller than that used for thinner samples (< 30 nm). Each diffraction pattern contains 256 × 256 pixels and is padded to 512 × 512 pixels for a better real-space sampling during reconstruction. Scan step size is 0.46 Å, probe forming semi-angle is 21.4 mrad, and the illuminating dose is $10^6$ e·Å$^{-2}$. (A) One selected diffraction pattern. (B) Position averaged diffraction pattern. (C) Total phase summing of all slices from ptychographic reconstruction. (D) Averaging amplitude of all slices from ptychographic reconstruction. The scale bar in (A) and (B) is 1 Å$^{-1}$ and the scale bar in (C) is 5 Å.

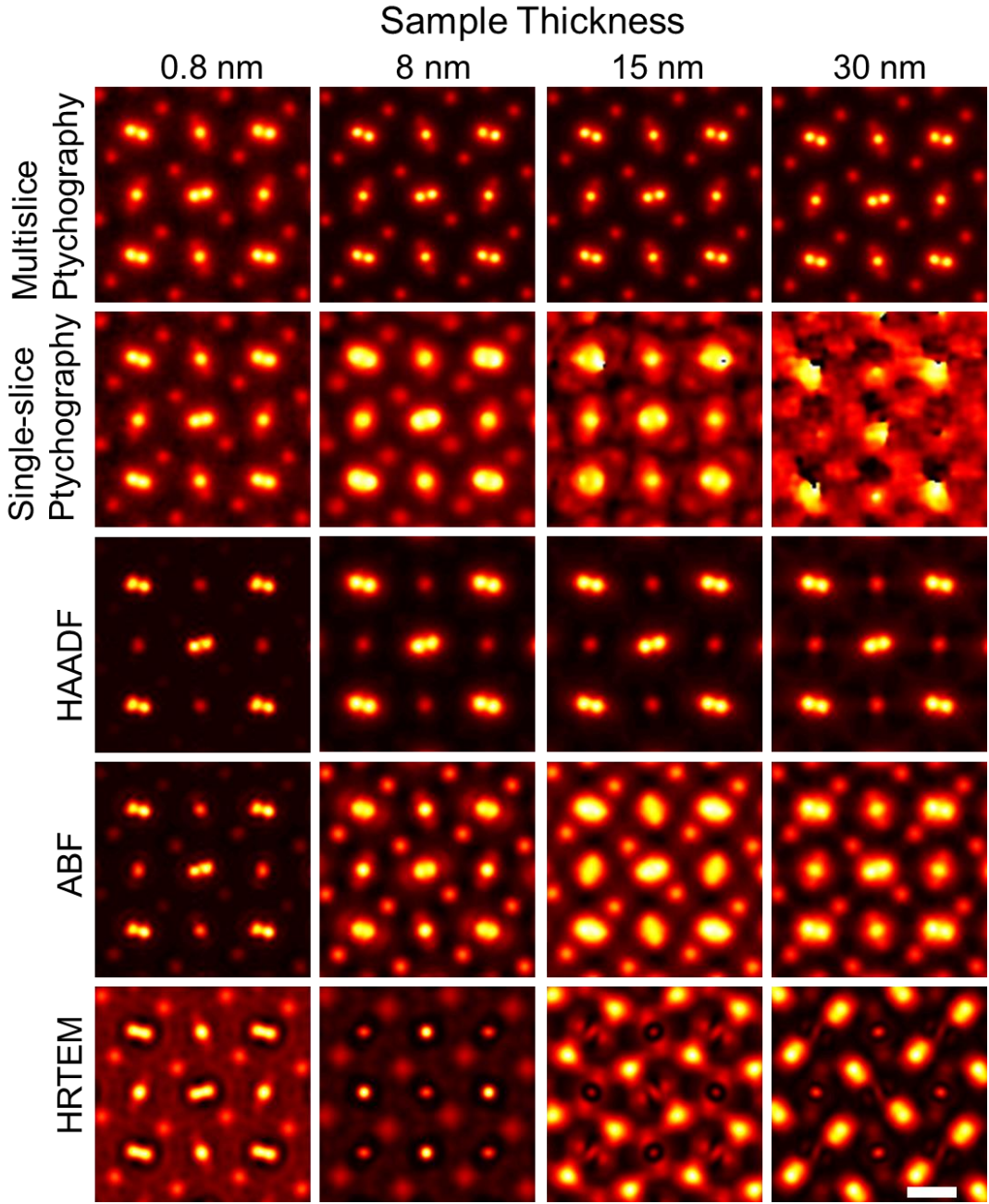

**Fig. S3.**

Simulated atomic-resolution images of PrScO$_3$ sample with different thicknesses for different imaging techniques. Probe forming semi-angle for ptychography, high angle annular dark-field (HAADF) and annular bright-field (ABF) images is 21.4 mrad. The contrast of the ABF images is inverted for direct comparisons with other image modes. Conventional high-resolution TEM (HRTEM) uses the optimal negative spherical aberration condition for imaging (*33*) with spherical aberration, $C_s$=-13 µm, defocus, 5.8 nm, and information limit, 1.25 Å$^{-1}$. The scale bar is 2 Å.

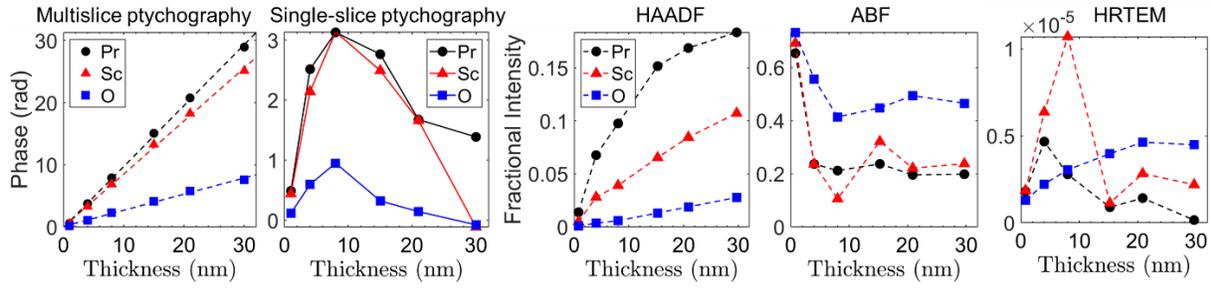

**Fig. S4.**

Intensity at atomic column positions of Pr, Sc and O from different images in fig. S3 as a function of thickness.

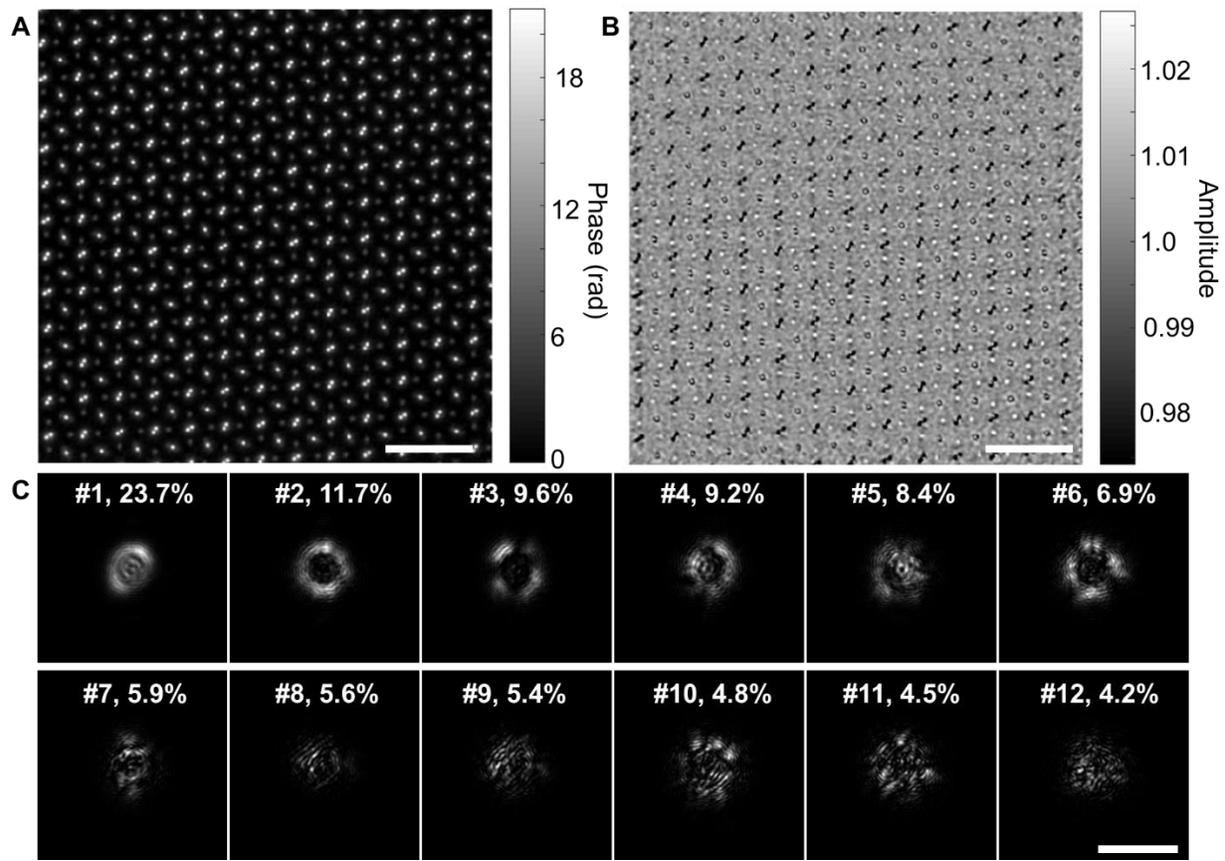

**Fig. S5.**

Phase (A) and amplitude (B) image of multi-slice electron ptychographic reconstruction from the whole experimental dataset. Intensity of all the probe modes from mixed-state algorithm are shown in (C). Indexes of modes and the corresponding fractional intensity in the total incident beam are labeled on the probe intensity images. The phase image is the summation of the phase of all object functions from different slices. The amplitude image is the average of the amplitude of all object functions from different slices. The scale bar is 1 nm.

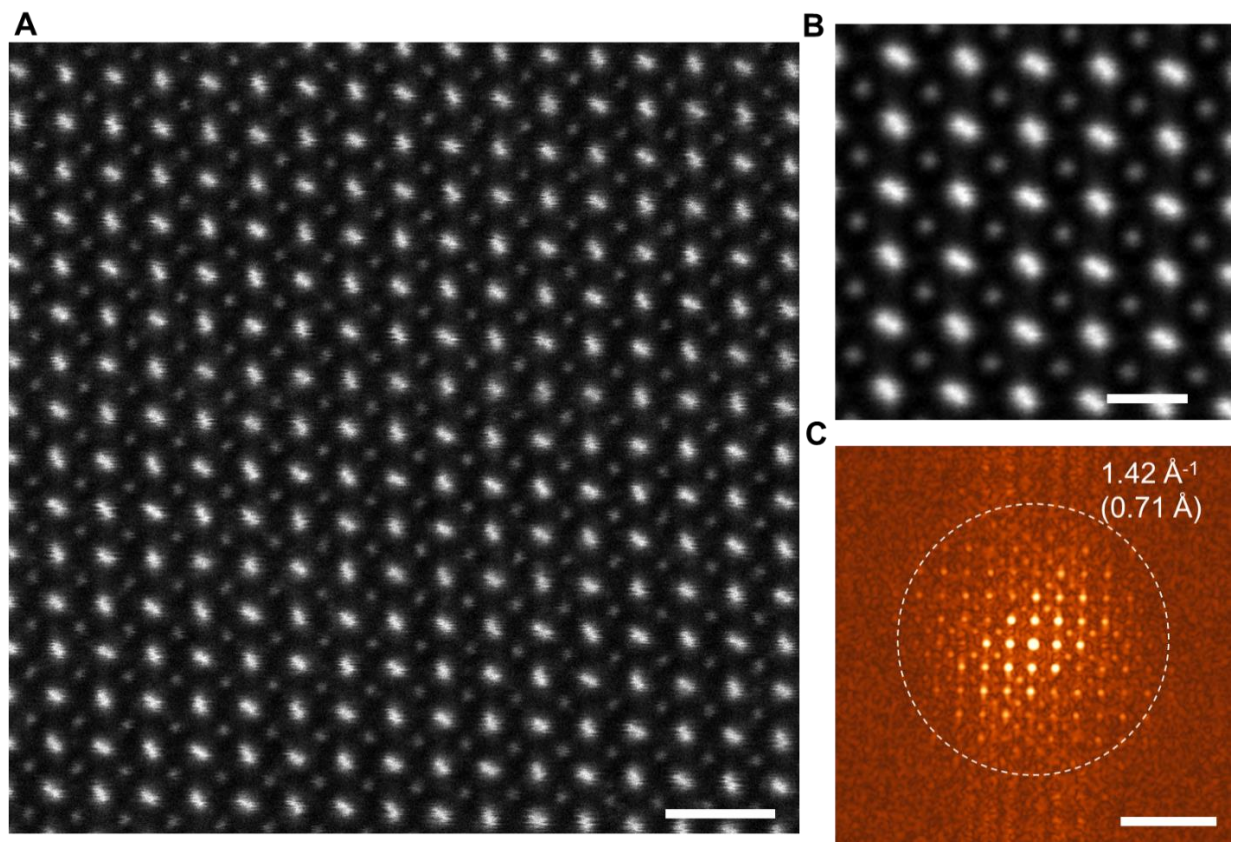

**Fig. S6.**

Image resolution from a conventional scanning transmission electron microscopy (STEM) high-angle annular dark-field (HAADF) image. (A) Single frame image of a large region. The scale bar is 1 nm. (B) Averaged image after image registrations using cropped regions from (A) as multiple frames to improve the signal-noise-ratio. The scale bar is 0.5 nm. (C) Fourier transformation of the image in (A). The resolution determined from (C) is 0.71 Å. The scale bar is 0.5 Å$^{-1}$.

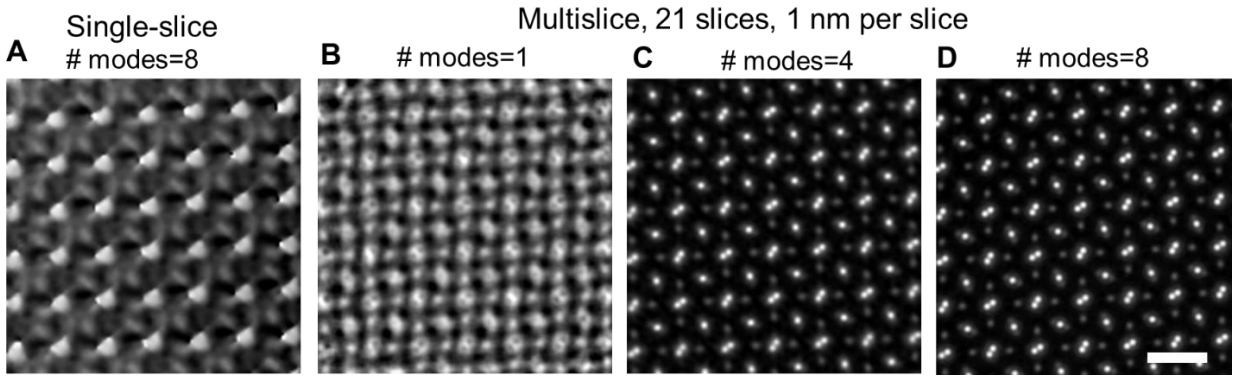

**Fig. S7.**

Reconstructed phase images from single-slice (A), multislice electron ptychography with single-state (B), and mixed-state (C) for four probe modes and (D) for eight probe modes for experimental dataset acquired from a 21-nm-thick $PrScO_3$ sample. The scale bar is 0.5 nm.

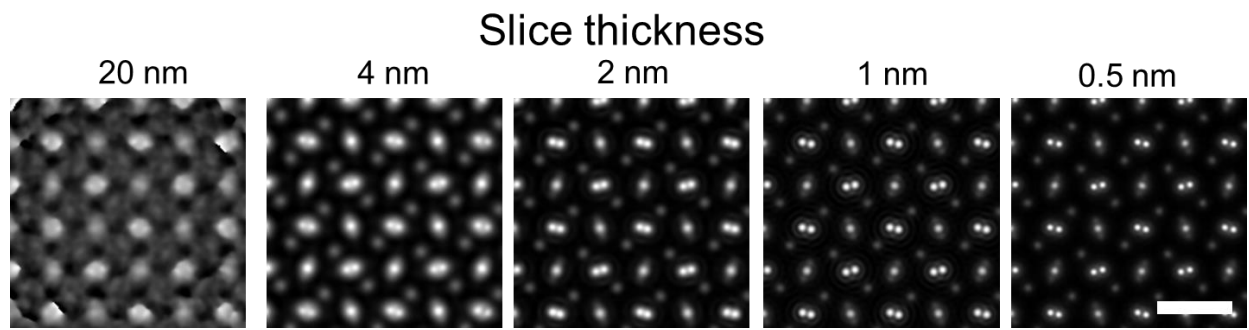

**Fig. S8.**

Reconstructed phase from multi-slice electron ptychography using different slice thicknesses. The data used is the simulated data from 20-nm-thick PrScO$_3$ sample. The scale bar is 0.5 nm.

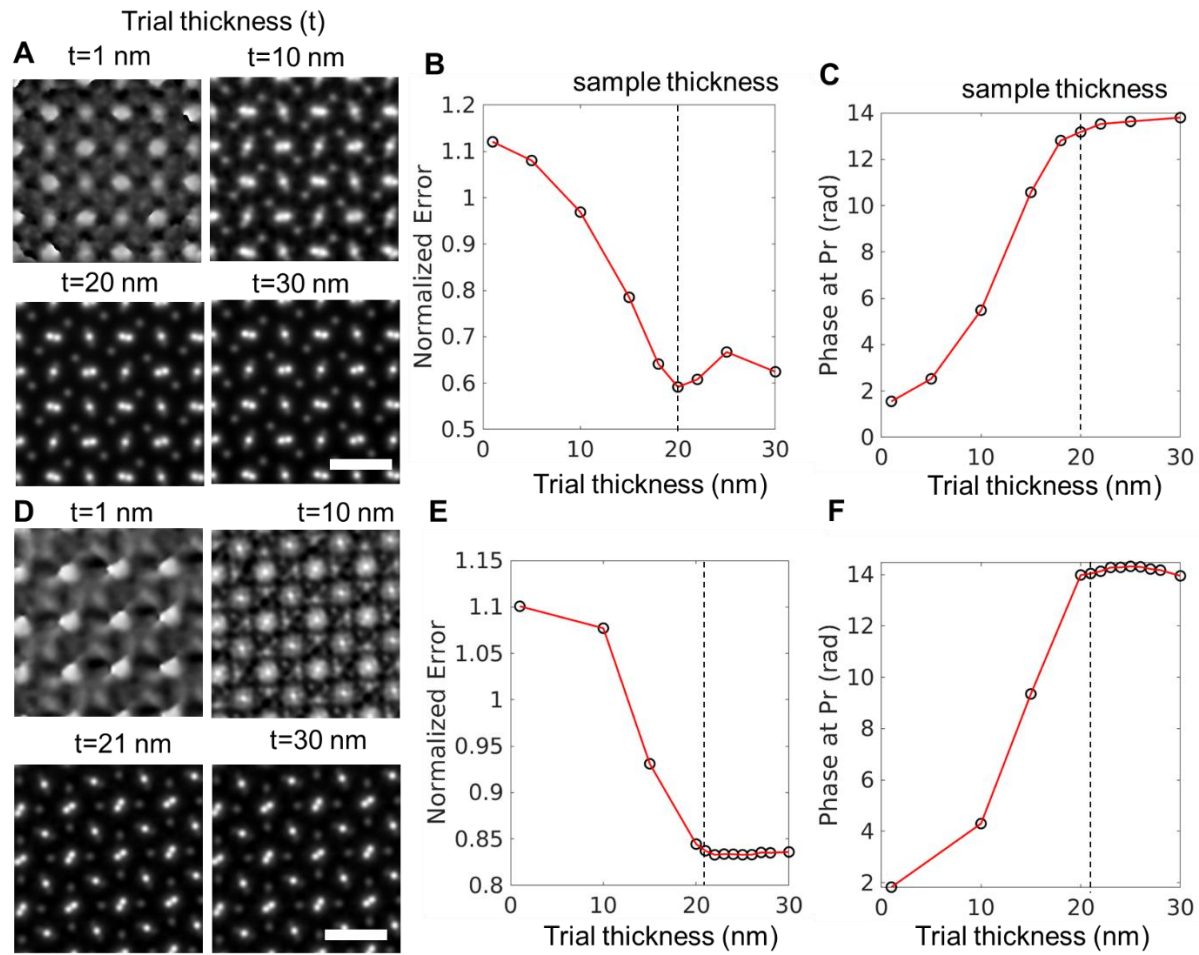

**Fig. S9.**

Determining the sample thickness from multi-slice electron ptychography. (A) Reconstructed phase images using different trial sample thicknesses for data simulated from a 20 nm thick $PrScO_3$ sample. (B) Normalized error from reconstructions using different trial thicknesses between the retrieved and experimental diffraction patterns. (C) The total phase change at Pr atomic site from reconstructions using different trial thicknesses. At the true sample thickness, the error shows a minimum and the total phase change reaches the maximum. (D) to (F) Similar as (A) to (C) but from experimental data. From the minimum of the error and the maximum of the total phase change, the thickness of the sample can be determined to be 21 nm. Thickness per slice is 1 nm during all reconstructions. The scale bar in (A) and (D) is 0.5 nm.

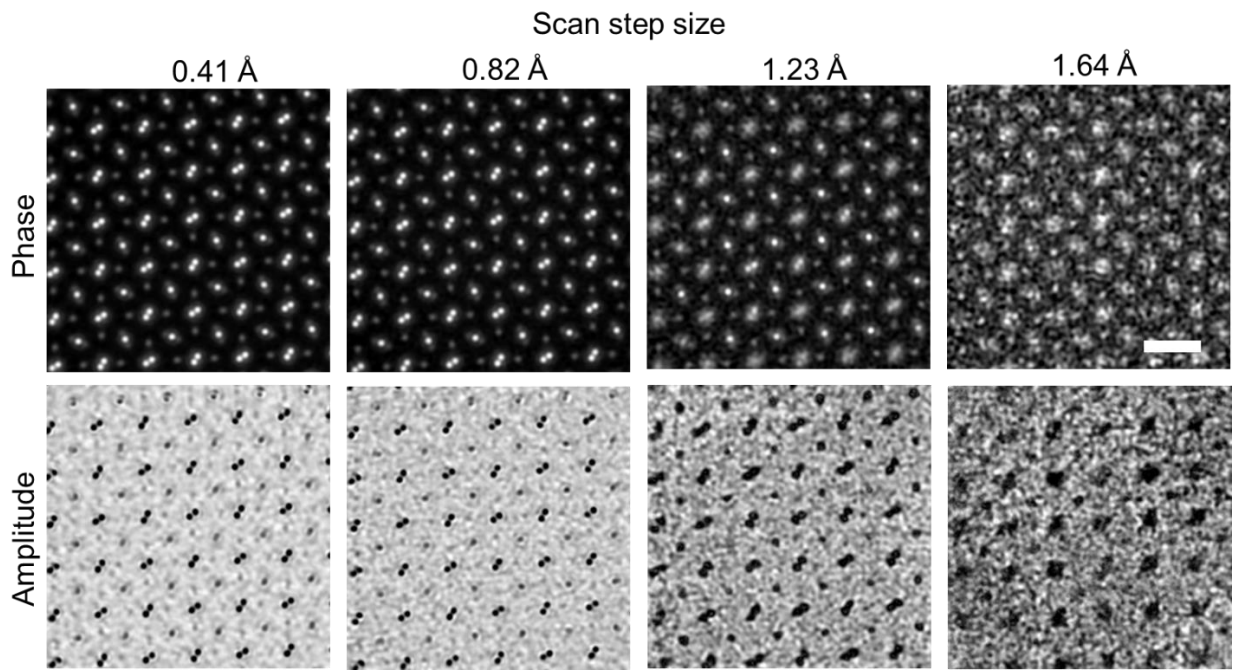

**Fig. S10.**

Reconstructed phase and amplitude from multi-slice electron ptychography using different scan step sizes from the experimental data. The sample is PrScO$_3$ with a thickness of 21 nm. The scale bar is 0.5 nm. Slice thickness of 1 nm was used during ptychographic reconstructions.

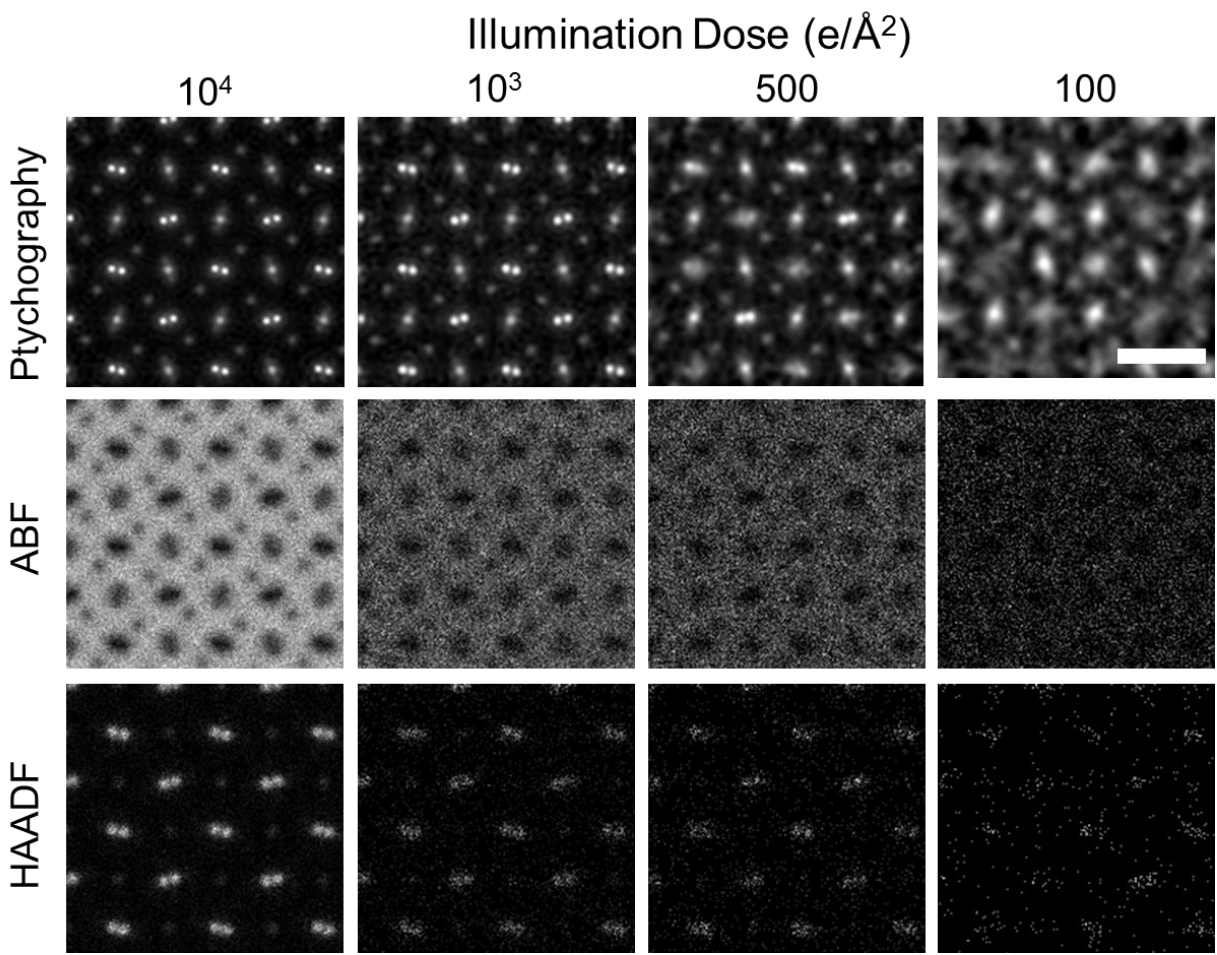

**Fig. S11.**

Simulations of multi-slice electron ptychography, annular bright-field (ABF) and high angle annular dark-field (HAADF) images at low dose conditions. Sample thickness is 15 nm. The scale bar is 0.5 nm.

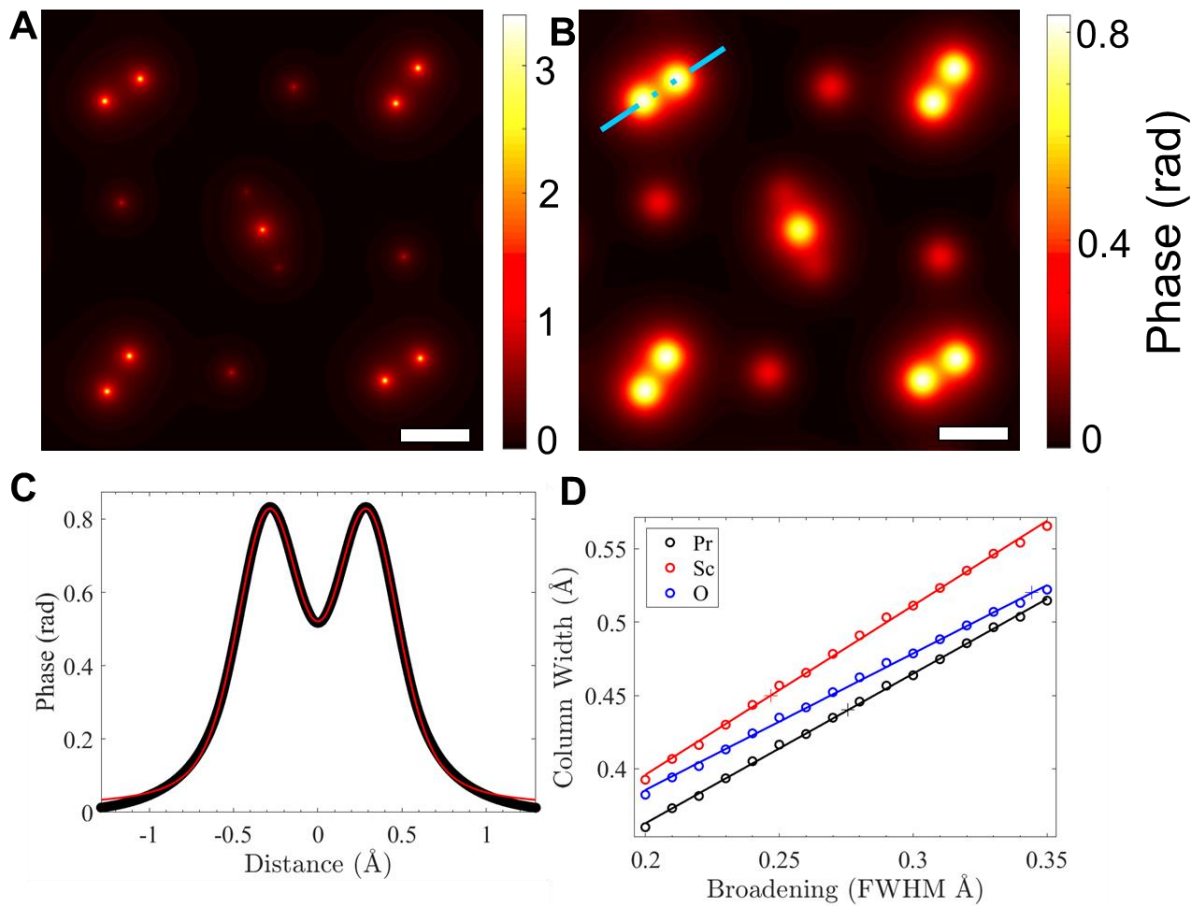

**Fig. S12.**

(A) Projected potential of PrScO$_3$ at 0 K. (B) One example potential after convolving a Gaussian function of the full-width half-maximum (FWHM), 0.28 Å. (C) One line profile along the Pr-Pr dumbbell (marked as blue line on (B)). The black dots are the data points, and the red line is the fitting of two Voigt functions. (D) The width of atomic columns in the potential broadened by Gaussian functions of different widths. The cross markers show the width of atomic columns from the experimental ptychographic reconstruction. The broadening factors (FWHM) for Pr, Sc, and O are 0.28 Å, 0.25 Å, and 0.34 Å, respectively. The scale bar in (A) and (B) is 1 Å.

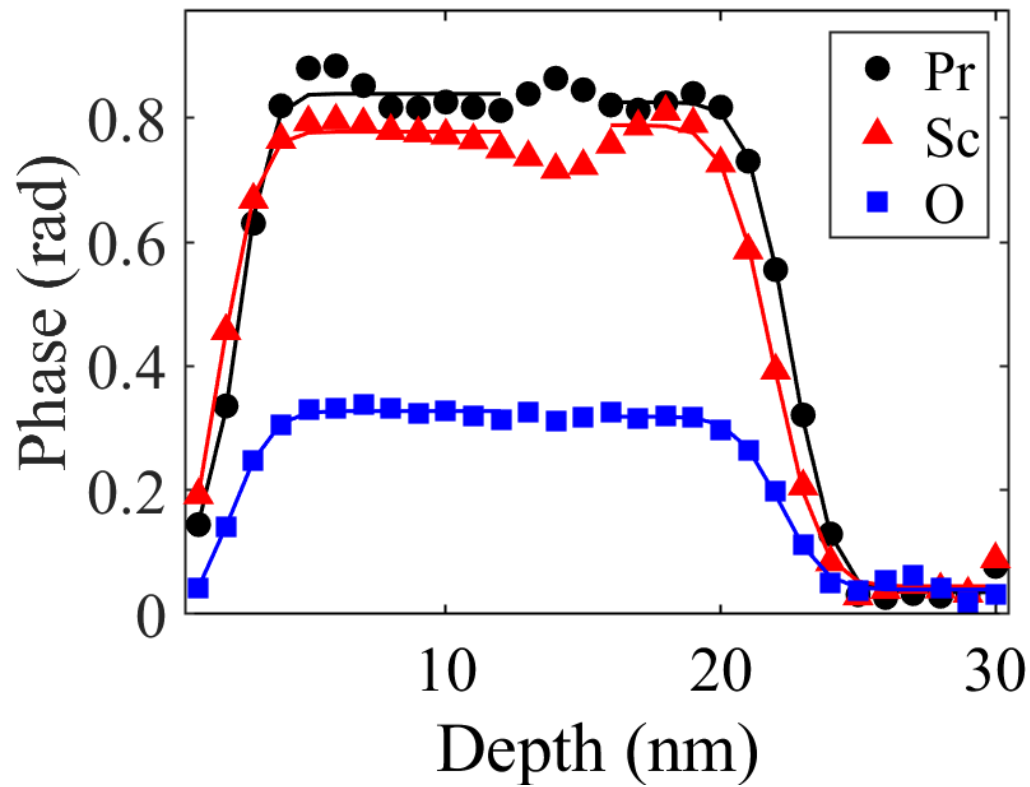

**Fig. S13.**

Depth profiles of phase at different atomic columns in ptychographic reconstruction from simulated data. Sample thickness is 20 nm and the dose is $10^6$ e·Å$^2$. Full width at 80 % of the maximum from Pr, Sc, and O positions is 2.0 Å, 2.3 Å, and 2.3 Å, respectively.

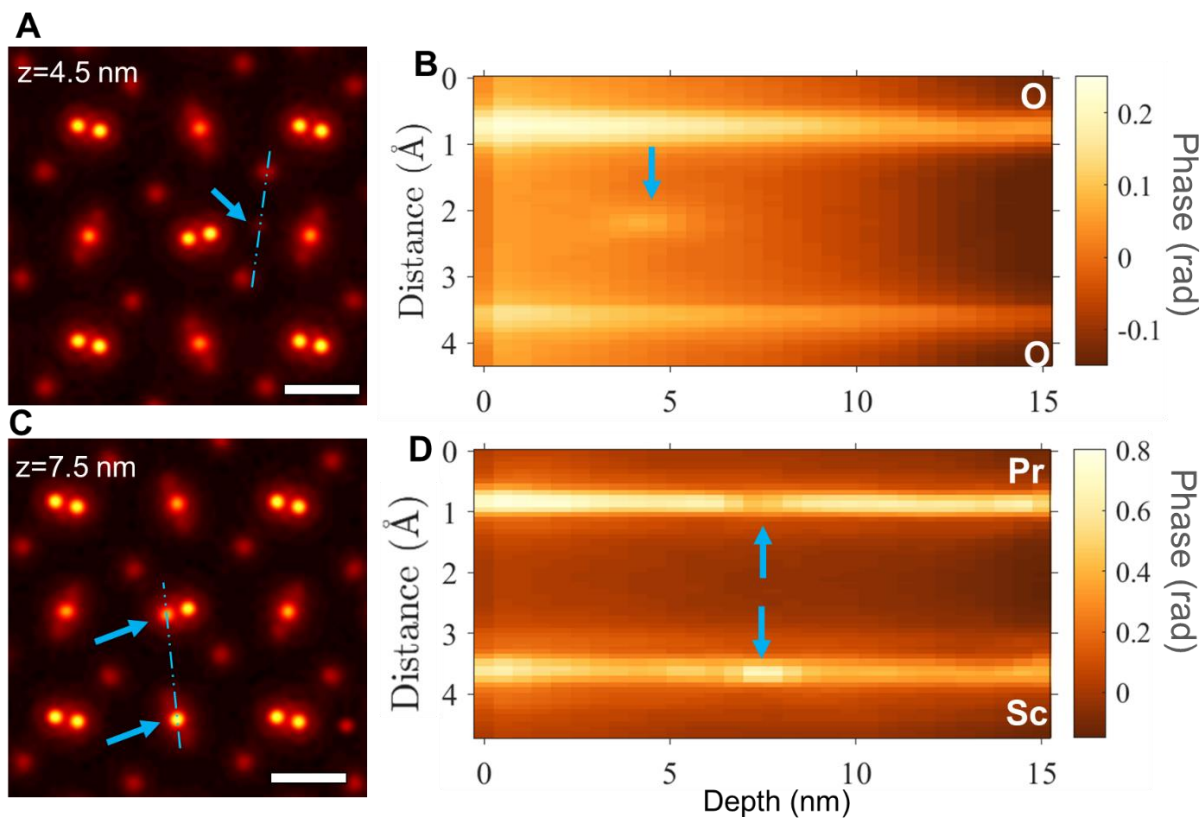

**Fig. S14.**

Depth profiles of phase across single Sc dopant and Pr/Sc anti-site dopants from simulations. (A) Phase image at depth, $z=4.5$ nm, showing the Sc dopant position. (B) Depth evolution across the direction marked on (A). (C) Phase image at depth, $z=7.5$ nm, showing the Pr/Sc anti-site dopant position. One Pr atomic site is replace by a Sc atom and one Sc atomic site is replaced by a Pr atom. (D) Depth evolution across the direction marked on (C). The illumination dose is $10^8$ e·Å$^{-2}$. Scale bar in (A) and (C) is 2 Å.

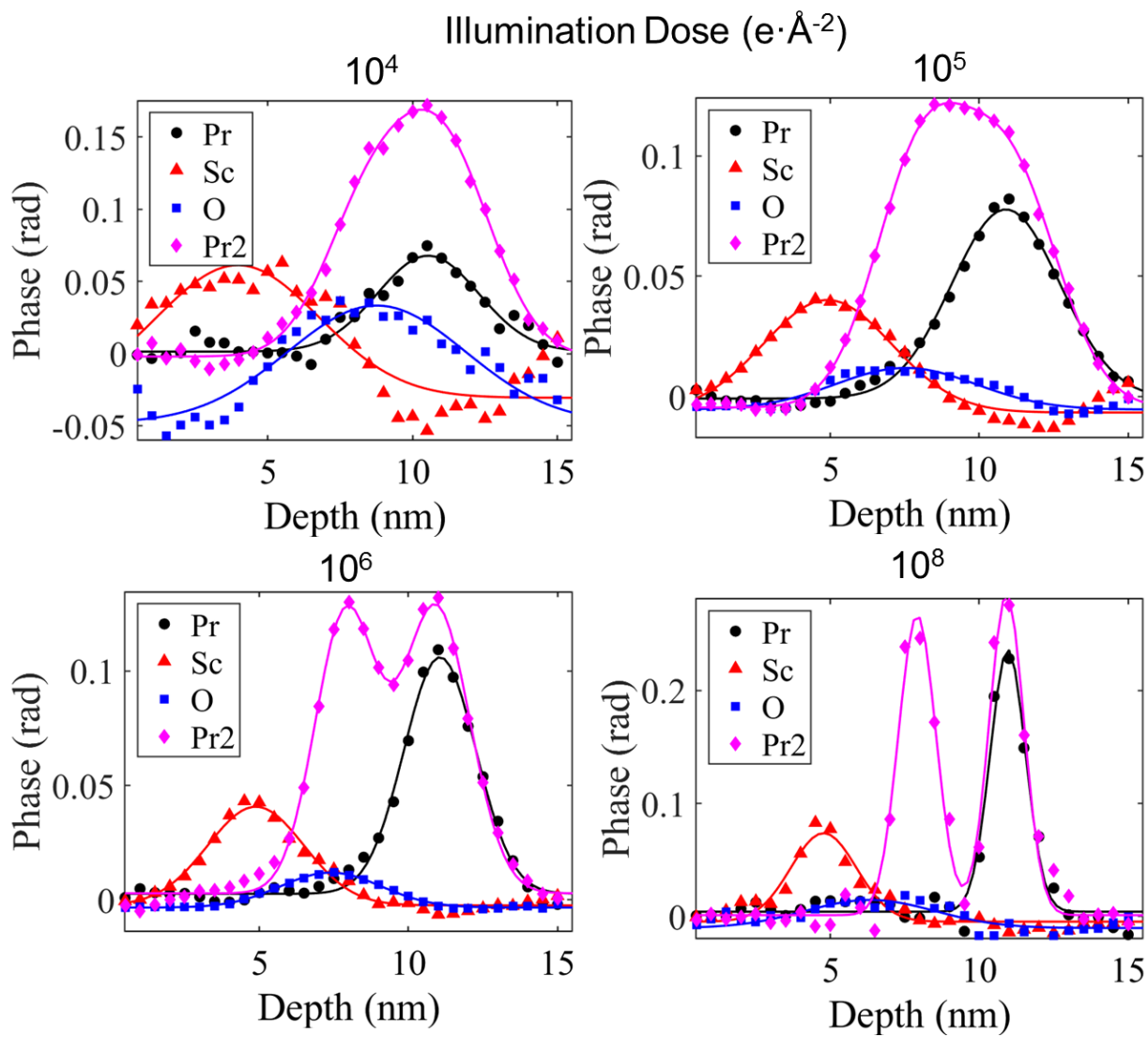

**Fig. S15.**

Depth profiles of the phase image from different dopants using different illumination doses by simulation. Better than 1.6 nm depth resolution (full-width at 80% of maximum) can be achieved at a $10^6$ e·Å$^{-2}$ dose from Pr atomic sites.

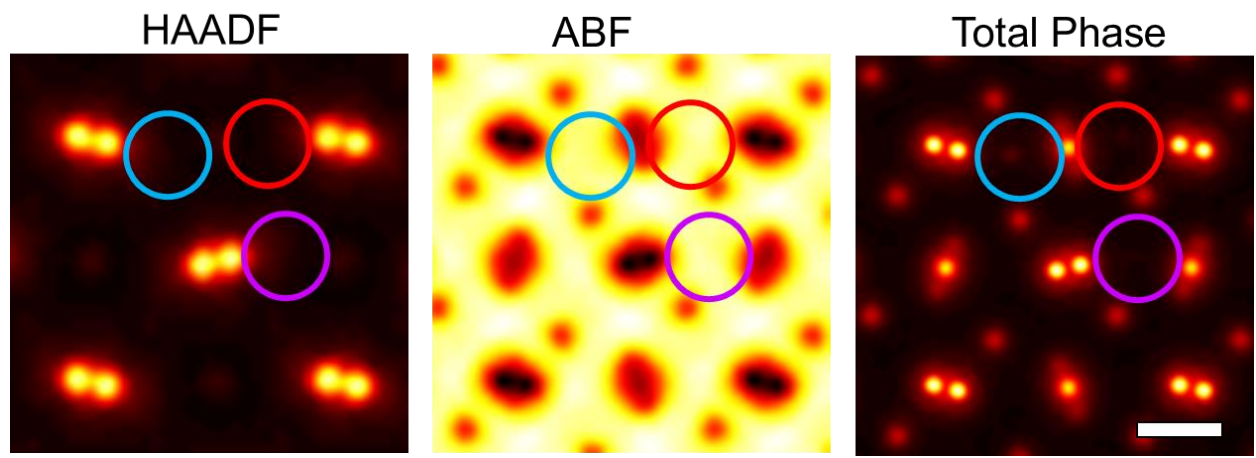

**Fig. S16.**

Dopant visibility from conventional projection imaging methods for the structural model used in Fig. 4. The blue, red, and purple circles mark the positions of the double Pr, single Pr and Sc dopant, respectively. For conventional high-angle annular dark-field (HAADF) and annular bright-field (ABF) imaging techniques, no contrast can be seen at any dopant site. In phase image summed up all layers from multislice electron ptychography, a very weak contrast shows up only at the double Pr site.  Contrast from single dopants is hidden by the strong tails of atoms from the matrix $PrScO_3$ structure. The scale bar is 2 Å.

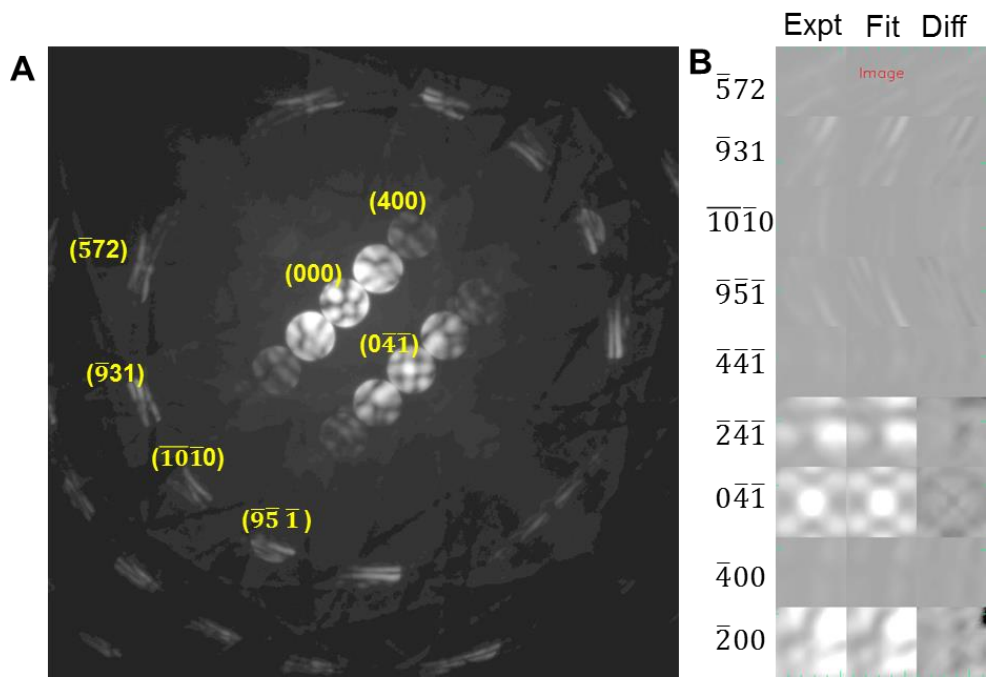

**Fig. S17.**

Debye-Waller factors determination of PrScO$_3$ using quantitative convergent beam electron diffraction (QCBED). (**A**) Experimental energy-filtered CBED pattern acquired with sample oriented ~10.6° away from the [001] zone axis, along the <010> mirror plane, through a 10-eV energy window around the zero-loss peak. High order Laue zone (HOLZ) Bragg disks sensitive to atomic positions and thermal vibrations were selected for refinement. (**B**) Debye-Waller factor refinement results by comparing the experimental diffracted intensities (Expt), calculated HOLZ lines (Fit), and their difference image (Diff). The isotropic Debye-Waller factors (*B*-factors) were determined as Pr: 0.737 Å$^2$, Sc: 0.595 Å$^2$, O#1 0.842 Å$^2$: O#2: 1.145 Å$^2$, in close agreement with reported single crystal X-ray diffraction results *(23)*. The CBED is indexed using lattice parameters a=5.6081 Å, b=5.7801 Å, and c=8.0252 Å, and spacegroup of No. 62.

|  | Different Atomic columns | | |
| --- | --- | --- | --- |
|  | Pr | Sc | O#2 |
| Column Width (Å) | 0.44 ± 0.01 | 0.45 ± 0.01 | 0.54 ± 0.02 |
| Total Broadening (Å) | 0.28 | 0.25 | 0.34 |
| Thermal Broadening (Å) | 0.23 | 0.20 | 0.24/0.28 |
| Abbe Resolution (Å) | 0.16 | 0.15 | 0.24/0.19 |
| Max. Phase (rad) | 22.4 ± 0.6 | 21.2 ± 0.3 | 6.7 ± 0.2 |
| Phase per Atom (rad) | 0.86 ± 0.02 | 0.41 ± 0.01 | 0.26 ± 0.01 |

**Table S1.**

Statistical analyses of the ptychographic reconstructed phase from the experimental data. Column width is the fitted full-width half-maximum (FWHM) of each atomic column using one Voigt function for Sc and O, and two for Pr columns. Total broadening is the FWHM of Gaussian determined from Fig. S12(D). Thermal broadening is the FWHM of thermal displacement from Debye-Waller factors ($B$ factor) determined from X-ray diffraction in (*23*) and QCBED. Abbe resolution is the FWHM of the broadening from ptychography, calculated from quadratically subtraction of the total broadening and thermal broadening. Thermal broadening factors and their corresponding Abbe resolutions of O#2 from both XRD and QCBED are listed. Max. phase is the maximum phase intensity at each atomic site. There are 26 unit-cells in the 21 nm-thick sample used in the experiment. In one unit-cell along the projection direction, there are only one Pr and O atom and two Sc atoms at each atomic column. The uncertainties of all quantities are the corresponding standard deviation.

| Elements | Debye-Waller B-factor (Å²) | | |
|---|---|---|---|
| | **Polycrystalline XRD (ref. (23))** | **Single crystal XRD (ref. (21))** | **QCBED (this study)** |
| Pr | 0.88 | 0.74 | 0.74 |
| Sc | 0.70 | 0.60 | 0.60 |
| O#1 | 1.00 | 0.84 | 0.84 |
| O#2 | 1.00 | 0.90 | 1.15 |

**Table S2.**

Debye-Waller factors (*B* factors) from two X-ray diffraction measurements and our quantitative convergent beam electron diffraction (QCBED).

**Movie S1.** Ptychographic reconstructed phase images of all slices from experimental data of $PrScO_3$ for main text Fig. 4A. The slice number is overlaid on the bottom-right corner of the images starting from the sample top surface. Each slice represents a thickness of 1 nm and there are 30 slices in total. The field-of-view of the region is $2.6 \times 2.6$ nm$^2$.

**Movie S2.** Ptychographic reconstructed phase images of all slices simulated from model structure with dopants for main text Fig. 4E. The slice number is overlaid on the bottom-right corner of the images starting from the sample top surface. Each slice represents a thickness of 0.5 nm and there are 30 slices in total. The field-of-view of the region is $1.6 \times 1.6$ nm$^2$. Sample thickness is 15 nm and the illuminated dose is $10^8$ e · Å$^{-2}$. The blue, red, purple, and yellow circles mark the positions of the double Pr, single Pr, Sc and O dopant, respectively.